\documentclass[aip,apl,12pt]{revtex4-2}
\usepackage{graphicx}
\usepackage{color}

\begin{document}

%\preprint{AIP/123-QED}

\title{3D arrangement of epitaxial graphene conformally grown on porousified crystalline SiC}

\author{S. Veronesi}
\email{stefano.veronesi@nano.cnr.it}
\affiliation{NEST, Istituto Nanoscienze-CNR and Scuola Normale Superiore, Piazza S. Silvestro 12, Pisa, 56127, Italy.}
\author{G. Pfusterschmied}%
\affiliation{Institute of Sensor and Actuator Systems, TU Wien, Vienna, 1040, Austria. 
}%

\author{F. Fabbri}
\affiliation{NEST, Istituto Nanoscienze-CNR and Scuola Normale Superiore, Piazza S. Silvestro 12, Pisa, 56127, Italy.}
\author{M. Leitgeb}%
\affiliation{Institute of Sensor and Actuator Systems, TU Wien, Vienna, 1040, Austria. 
}%
\author{O. Arif}
\affiliation{NEST, Istituto Nanoscienze-CNR and Scuola Normale Superiore, Piazza S. Silvestro 12, Pisa, 56127, Italy.}
\author{D.A. Arenas}
\affiliation{EMAT, University of Antwerp, Groenenborgerlaan 171, Antwerp, B-2020, Belgium.}
\author{S. Bals}
\affiliation{EMAT, University of Antwerp, Groenenborgerlaan 171, Antwerp, B-2020, Belgium.}
\affiliation{Nanolab Centre of Excellence, Groenenborgerlaan 171, Antwerp, B-2020, Belgium.}
\author{U. Schmid}%
\affiliation{Institute of Sensor and Actuator Systems, TU Wien, Vienna, 1040, Austria. 
}%
\author{S. Heun}
\affiliation{NEST, Istituto Nanoscienze-CNR and Scuola Normale Superiore, Piazza S. Silvestro 12, Pisa, 56127, Italy.}

\date{\today}% It is always \today, today,

\begin{abstract}
Nanoporous  materials represent a versatile solution for a number of applications ranging from sensing, energy applications, catalysis, drug delivery, and many others. The synergy  between the outstanding properties  of graphene with a three-dimensional porous structure, circumventing the limits of its 2D nature, constitutes therefore a breakthrough for many fields. We report the first three-dimensional growth of epitaxial graphene on a porousified crystalline 4H-SiC(0001). The wafer porosification is performed via a sequence of metal-assisted photochemical and photoelectrochemical etching in hydrofluoric acid based electrolytes. Pore dimensions of the matrix have been evaluated by electron tomography resulting in an average diameter of 180 nm. Graphene growth is performed in an ultra high vacuum environment at a base pressure of $10^{-11}$ mbar. The graphene growth inside the pores is uniform as confirmed by Transmission Electron Microscopy (TEM) analysis. Raman spectroscopy confirms the high quality of the graphene with a 2D/G ratio $>1$ and an average graphene crystal size of $\approx$ 100 nm. Furthermore, it demonstrates a uniform coverage of graphene across the whole sample area. The surface-to-volume ratio of this novel material, its properties, the tunability of the pore size and the scalability of the surface porosification process offer a game changing perspective for a large number of applications.%
\end{abstract}

\keywords{}%Use showkeys class option if keyword
\maketitle

\section{Introduction}

Graphene's outstanding properties make it an ideal material to realize high performance devices and sensors. Its huge potential has driven a scientific revolution, first using pristine graphene for optoelectronic applications \cite{Wang}, flexible electronics \cite{Shrivas, Singh}, graphene-based sensors \cite{Singh, Nag, Kumar, WangC}, and biological applications \cite{Merkoci}. Large research efforts have been made to tailor the electronic properties of graphene through defect engineering \cite{Bouk, Ortiz, Ye} and chemical functionalization, in order to increase the application fields and to improve the performance of devices. However, the 2D nature of graphene represents a limit for many applications: catalysis \cite{Yan2020}, photoassisted water splitting \cite{Sun}, gas detection \cite{wang2015} and storage \cite{Mohan2019, Jain2019}, drug delivery \cite{McCallion}, high performance electrodes \cite{Tiliakos, Thimmappa}, supercapacitors \cite{Venkateshalu, Ping17}, battery cathode \cite{Shen2021}, water treatment and filtration \cite{Safarpour, Ray}, all would strongly benefit from a high surface-to-volume ratio and a 3D structure. 

The technological importance of porous carbon materials is underlined by a vast literature which starts even before the graphene discovery \cite{Han98}, in the form of carbon foams which represent an economical solution. Such carbon foams are prepared by different routes: the pyrolysis of carbon precursors, the assembly of exfoliated graphite or graphene oxide, always with chemical additives \cite{Inagaki15}. Porous carbon materials are extensively utilized for latent heat thermal storage \cite{Zhong10}, for adsorption of large molecules \cite{Rajib}, as thermal conductive materials \cite{Gallego03}, for electrodes \cite{Li17, Xue} and for supercapacitors\cite{Inagaki15, Ping17}. However, the material quality is much lower than epitaxial graphene. Cleaner, high-performance materials are graphene foams grown by chemical route or chemical vapour deposition (CVD) \cite{Teich, Zhang16}, and porous graphene networks\cite{D0MH00815J} which represent a first step toward enhanced material quality, but are still far from the quality of epitaxial graphene. A three-dimensional arrangement of pillared graphene for hydrogen storage \cite{Dimitrakakis} and mass sensor applications \cite{Duan} has been theoretically investigated but never realized. Therefore, the possibility to achieve a 3D graphene assembly without losing its outstanding 2D properties is an ongoing research challenge. 

Here we propose a game changing approach that allows to overcome these technological bottlenecks. It is based on the idea to grow an epitaxial graphene layer on a three-dimensional backbone constituted by nano-porous crystalline SiC \cite{Leitgeb}. The epitaxial graphene is grown by thermal decomposition of the SiC in an ultra-high vacuum (UHV) environment, resulting in a native 3D graphene material with a light, but overall robust structure. This is to the best of the authors` knowledge the first approach for the realization of 3D epitaxial graphene, a novel idea which enables a new and versatile material. 

Porous SiC (pSiC) \cite{Shor1993} has been initially developed for the fabrication of SiC MEMS devices based on a novel surface micromachining process \cite{Leitgeb2017a} and successively utilized to fabricate optical components, such as rugate mirrors \cite{Leitgeb2017} fully integrated into a 4H-SiC substrate. The fabrication process has already been optimized and allows to control the local definition of the pores as well as the degree of porosity with depth \cite{Leitgeb1}. Moreover, the porosification technique allows obtaining stacked layers of different porosity, increasing the versatility of this material \cite{Leitgeb3}. Commercial wafers of 4H-SiC(0001) and (000$\bar{1}$) have been porousified to produce samples of porous SiC in both Si-face and C-face orientation. This technique is scalable to large area samples \cite{Leitgeb2}. Furthermore, porousified areas can be arranged in arrays, and it is possible to micromachine cavities in between porousified and dense 4H-SiC \cite{Leitgeb2017a}. This versatility, and the coupling of a 3D graphene structure to a semiconductor backbone, opens up the possibility for a large number of applications in nano-electronics, sensors, and energy related fields.

\section{Materials and Methods}
\begin{figure*}
   \begin{center}
   \includegraphics[width=0.9\textwidth]{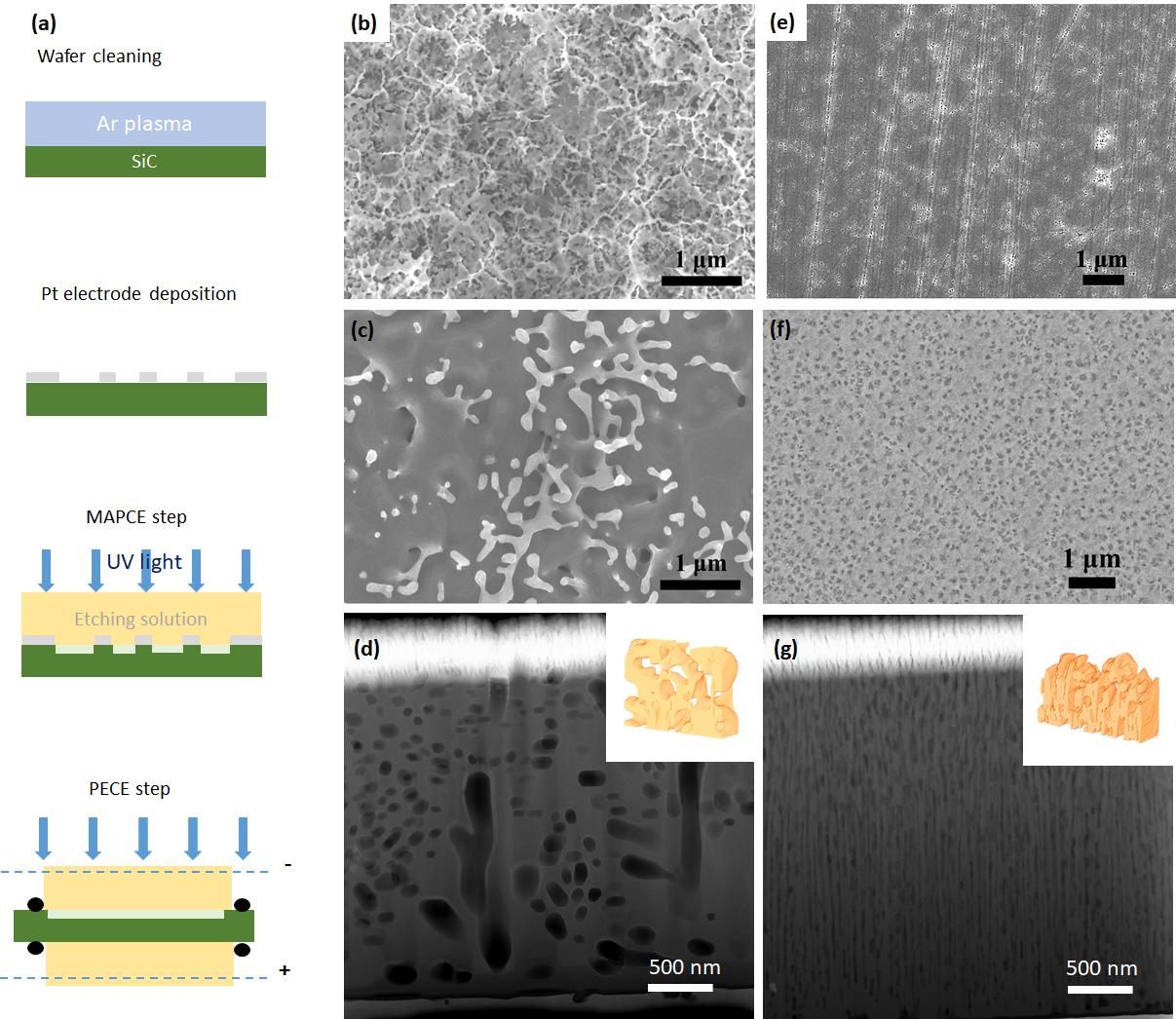}
   \end{center}
\caption{\label{SEM} (a) Scheme of the MAPCE-PECE porousification process. (b) Plan-view SEM images of a Si-face sample from a 4H-SiC wafer after the porousification and (c) after the UHV annealing at $1370^\circ$~C. (d) Cross-sectional TEM image of a Si-face sample, after the UHV annealing at $1370^\circ$~C, prepared by FIB; the inset shows the electron tomography reconstruction of the pore structure. (e) Plan-view SEM images of a C-face sample from a 4H-SiC wafer after the porousification and (f) after the UHV annealing at $1260^\circ$~C. (g) Cross-sectional TEM image of a C-face sample, after the UHV annealing at $1260^\circ$~C, prepared by FIB; the inset shows the electron tomography reconstruction of the pores structure.}
\end{figure*}

\textbf{Porous SiC} (Si-face) is obtained by utilizing a metal-assisted photochemical etching (MAPCE) followed by photo-electrochemical etching (PECE) of 4H-SiC \cite{Leitgeb2}, as shown schematically in Figure \ref{SEM}(a). The MAPCE step is introduced to produce a porous layer in the $\mu$m range \cite{Leitgeb2} and provides the initialization sites for the PECE process.
To perform MAPCE step, 300~nm thick Pt pads are deposited on the wafer, which act as local cathode and low-resistance charge transfer to the electrolyte. An etching solution containing Na$_2$S$_2$O$_8$ (0.15~mol/L) and HF (1.31~mol/L) is then used for MAPCE, performed in a standard electro-chemical cell (AMMT GmbH). A 250~W ES280LL mercury lamp at full spectrum was used as UV source in both MAPCE and PECE steps.
 After MAPCE, a PECE \cite{Leitgeb2} step is performed. The porousification of C-face samples is obtained with a single PECE step. During PECE, a solution containing HF (5.52~mol/L) and ethanol (1.7~mol/L) is utilized, while UV exposure is running.

Porous SiC samples were analyzed via Scanning Electron Microscopy (SEM) (see Figure \ref{SEM}(b) and \ref{SEM}(e)), TEM and electron tomography (see Figure \ref{SEM}(d) and \ref{SEM}(g)), and Raman spectroscopy (see Figure S5). SEM images of the samples were obtained with a Zeiss Merlin SEM operated at 5~keV. For each sample, the surface morphology was investigated from plan-view SEM micrographs taken from at least ten different areas of the sample. SEM images of the porous 4H-SiC Si-face sample show a very rough surface with a "dendritic" structure (see Figure \ref{SEM}(b)), produced by the MAPCE step. The C-face sample shows a different morphology. The surface is almost flat, due to the absence of the MAPCE passage, with craters well visible in Figure \ref{SEM}(e). Raman spectroscopy and mapping was performed by using a Renishaw confocal microscope with a $100 \times$ objective (NA 0.85) equipped with a 532~nm laser as excitation source. The laser power is 1~mW on a spot size of 1~$\mu$m. The scan area is $20 \times 20$~$\mathrm{\mu m^2}$, with 400 pixels, and a lateral pixel size of 1~$\mu$m.

\textbf{The graphene growth} was performed by thermal decomposition of the SiC, annealing the samples in an UHV environment with a base pressure of $5 \times 10^{-11}$ mbar. It was performed by direct current heating with a Keithley 2260B power supply, while temperature was monitored with a J.S.C. TSS-F1-C3000 optical pyrometer. Once the target temperature had been reached, it was kept constant for 2 minutes before cooling down to room temperature. The optimal growth temperature and annealing time depend on the use of the C- or Si-face of the wafer. For both the Si- and C-face, up to about $1140^\circ$~C graphene does not grow, and Raman spectra are virtually identical to those of the starting pSiC sample. For Si-face samples, graphene starts to show a Raman signature after an annealing at temperatures higher than $1320^\circ$~C and reaches its highest quality at $1370^\circ$~C. At even higher temperatures the quality remains constant up to $1470^\circ$~C, the highest explored temperature. As expected, the growth window for C-face samples is much narrower. Graphene starts to grow at $1220^\circ$~C, reaches its best quality at $1260^\circ$~C, but becomes graphitic already at $1300^\circ$~C. 

On the Si-face samples, graphene growth by thermal decomposition produces a surface reordering due to Si evaporation from the tiny branches of the dendritic structure, as clearly shown in Figure \ref{SEM}(c) and S1 (a,b). After the graphene growth on the C-face sample, the surface does not change much, and reordering is less evident, as shown in Figure \ref{SEM}(f) and S1 (c,d). 

The pore structure in a sample section has been investigated by preparing proper lamellas by Focused Ion Beam (FIB) and analyzing them by electron tomography in High--Angle Annular Dark Field - Scanning Transmission Electron Microscopy (HAADF-STEM) mode for the 3-D characterization of the porous SiC. The analysis has been performed using a Thermo Fisher Scientific Titan microscope, equipped with an aberration corrector and a monochromator. The ability of the microscope to work at different voltages (80~kV to 300~kV) and the presence of a monochromator provides enough flexibility to control and minimize electron beam damage to the sample for the tilt series acquisition as well as for the visualization of the graphene layers at high resolution, discussed in the following. HAADF electron tomography tilt series were acquired by using a Fischione 2020 tomography holder between $-75^\circ$ and $75^\circ$ with a step of $3^\circ$. The projection images were acquired at a magnification of 80~k$\times$ and image resolution of $2048 \times 2048$ pixels. The obtained series were aligned using cross-correlation, and 3D reconstructions were obtained using the Expectation Maximization (EM) algorithm, as implemented in Astra Toolbox \cite{vanAarle2015}. The reordering of the 3D pore structure is shown in Figures \ref{SEM}(d) and (g). Pore dimensions of the Si-face sample, obtained by electron tomography and shown in the inset to Figure \ref{SEM}(d), are peaked at around 80~nm in diameter, with an average of 180~nm (see Figure S2). The pores occupy about 33\% of the sample volume. 
In the C-face sample, pores are smaller, see Figure \ref{SEM}(g), and their diameter peaks between 30 and 70~nm, with an average of 85~nm. The pores occupy about 42\% of the sample volume. Further details are given in the Supporting Information.

\section{Results and discussion}
%\label{}

\begin{figure*}
   \begin{center}
   \includegraphics[width=0.9\textwidth]{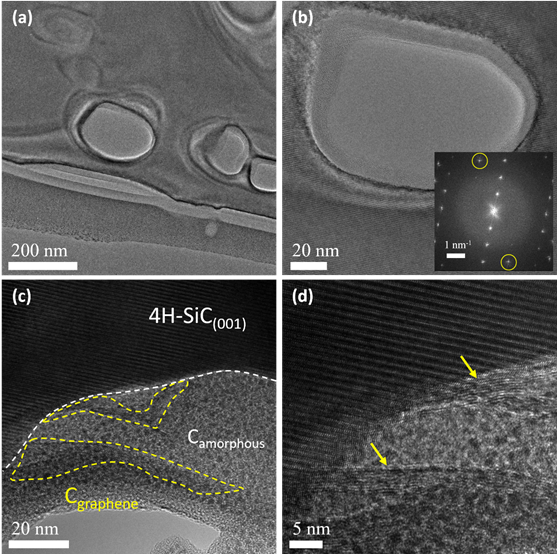}
   \end{center}
\caption{\label{TEM} High resolution TEM images of a Si-face sample after graphene growth at $1370^\circ$~C. (a) TEM image at 300 kV from an area 826~nm $\times$ 826~nm showing multiple pores, (b) magnification to 152~nm $\times$ 152~nm on one of the pores, inset FFT analysis with two extra spots of possible (002) graphene spacing encircled in yellow. (c) TEM image at 80 kV from an area of 81~nm $\times$ 81~nm showing stacked graphene layers, indicated by the yellow ellipses, (d) 31~nm $\times$ 31~nm on the left-central area in panel (c) where the stacked graphene layers are marked by yellow arrows.}
\end{figure*}

\textbf{Si-face samples.} 
To demonstrate that graphene growth was successful in the 3D pores of the porous material, the FIB technique has been utilized to obtain lamellas from the Si-face sample. Observing the sample at 300 kV in TEM, the roundish pore morphology is evident in Figure \ref{TEM}(a) and (b). The FFT analysis of Figure \ref{TEM}(b) shows the (001)~4H-SiC array with points spaced by 1.01~nm, with two extra spots encircled in yellow (see inset on Figure \ref{TEM}(b)). These extra spots do not belong to the dominant crystal structure and correspond to an interplanar distance of 0.34~nm, consistent with the graphene inter-layer distance, which could indicate a stacking of the graphene sheets inside pores. To confirm this hypothesis, TEM analysis at lower electron energies (80~keV) has been performed, which avoids sample degradation due to the electron beam and allows to observe the graphene signature inside the pores. Figure \ref{TEM}(c) shows a higher resolution image of one of the pores where the SiC crystal structure can be seen at the outer part of the hole while the inner part appears to be composed of amorphous carbon, where it is possible to observe graphene layer stacking, especially in the higher magnification image on Figure \ref{TEM}(d) marked by yellow arrows, with an interspace distance of about 0.34~nm, in agreement with literature values for graphite. 

\begin{figure*}
   \begin{center}
   \includegraphics[width=0.9\textwidth]{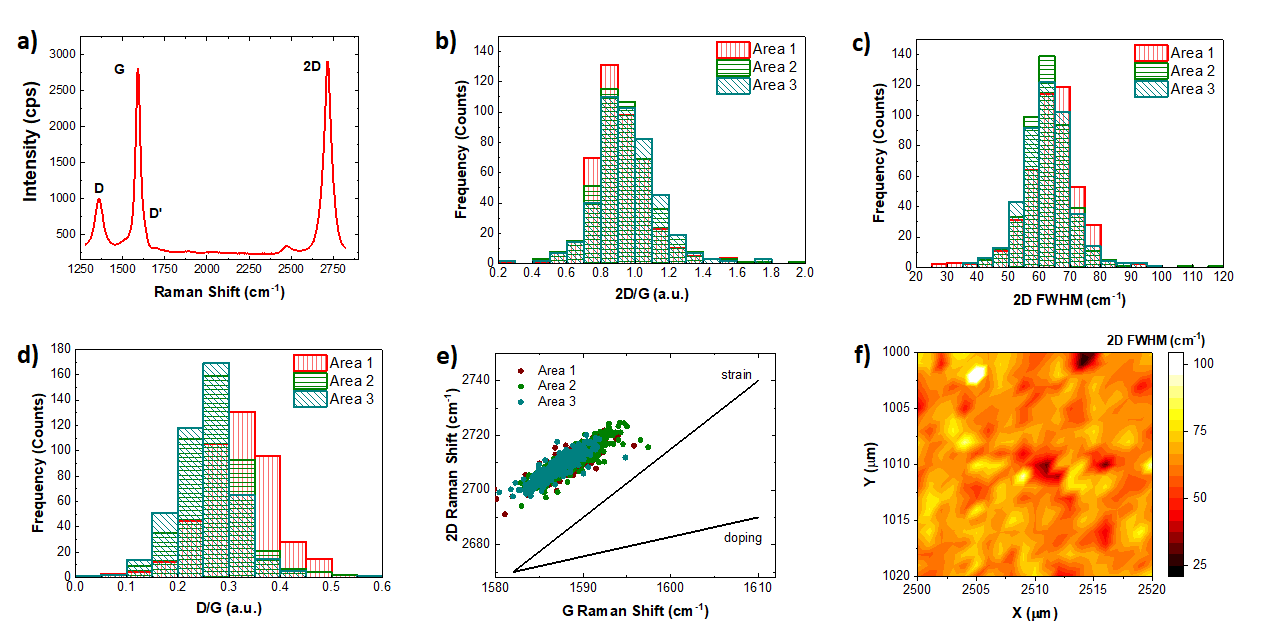}
   \end{center}
\caption{\label{raman1} Raman analysis of a porous 4H-SiC Si-face sample after annealing at $1370^\circ$~C. (a) Raman spectrum from 1250 to 2850~cm$^{-1}$ (graphene peaks), (b) superposition of the 2D/G ratio histograms of three different areas on the sample, (c) analysis of the FWHM width of the 2D peak of the areas 1, 2 and 3 (superposition of the three histograms), (d) superposition of the D/G ratio histograms of the areas 1, 2 and 3, (e) Raman peak shift analysis, (f) map of the 2D peak width relative to area 1 as an example, full maps in SI.} 
\end{figure*}

Raman analysis shown in Figure \ref{raman1}(a) consistently confirms the successful synthesis of graphene on the porousified Si-face 4H-SiC substrate. The spectrum presents the three prominent Raman modes of graphene, the D peak (at 1359~cm$^{-1}$), the 2D peak (at 2716~cm$^{-1}$), and the G peak (at 1592~cm$^{-1}$) \cite{Ferrari06, Malard09}. We observed also the presence of a shoulder to the G peak due to the D’ defect activated Raman mode \cite{Eckmann12}. The 2D peak has a single Lorentzian shape and a full width at half-maximum (FWHM) of $\approx 55$~cm$^{-1}$, a signature of single-layer graphene \cite{Ferrari06}.

In Figure \ref{raman1}(a), we underline the complete absence  of the second order SiC-related Raman modes peaked at 1516~cm$^{-1}$ and the two-phonon band at 1750~cm$^{-1}$ \cite{Hahnlein}. In order to clarify the absence of these peaks, we analyze the range of Raman shifts related to the transverse optical (TO) phonons of 4H-SiC, discussed further on. 

In order to evaluate the homogeneity of the graphene film, scanning Raman mapping has been carried out in 3 areas of the sample, with a size of 20~$\mu$m~$\times$~20~$\mu$m  for each area (see Figures S6 - S10). The statistical analysis of the standard benchmarks of graphene, such as the 2D/G intensity ratio, the 2D FWHM, and the D/G intensity ratio, obtained in the different areas, are presented in Figure \ref{raman1}(b), \ref{raman1}(c) and \ref{raman1}(d), respectively. 

The histogram of the 2D/G intensity ratio, reported in Figure \ref{raman1}(b), shows a good homogeneity over the three different areas. In particular, the average values of the 2D/G intensity ratios are 0.90~$\pm$~0.10, 0.92~$\pm$~0.10, and 0.94~$\pm$~0.10 for area 1, area 2 and area 3, respectively. The error associated to the average value is the FWHM of the Lorentzian curve, used to fit the histogram. The 2D/G ratio is normally employed for the evaluation of the number of layers in graphene, but it is also strongly sensitive to the graphene doping. Therefore, considering the perfect Lorentzian fit of the 2D peak, indication of monolayer graphene, it is possible to deduce a high doping level of the graphene obtained on the porousified 4H-SiC substrate. Considering the 2D/G ratio around 1, we can hypothesize a doping concentration of $2 \times 10^{13}$~cm$^{-2}$ \cite{Ferrari09,Das}. A high electron doping of epitaxial graphene is well known and reported to be due to charge transfer from the substrate \cite{Lee}.

Another important benchmark for the quality of graphene is the width of the 2D peak, used for the evaluation of the electron-electron scattering as well as the strain. This is important due to the complex morphology of the substrate on which the graphene is synthetized. The average FWHM of the 2D peak in the different areas is 64.0~$\pm$~5.0~cm$^{-1}$ (area 1) and 62.5~$\pm$~5.0~cm$^{-1}$ (areas 2 and 3). In Figure \ref{raman1}(f) the 2D FWHM of area 1 is reported as an example. For epitaxial monolayer graphene obtained on flat 4H-SiC, the  FWHM of the 2D peak is around 50~cm$^{-1}$ \cite{Lee}. The increase of the FWHM indicates a strain modification for the graphene obtained on the porousified Si-face 4H-SiC substrate \cite{Neumann}. 

The last standard benchmark for the graphene quality is the D/G intensity ratio. This parameter is important for both the defect nature \cite{Eckmann12} and their concentration \cite{Canado11} and the grain size in polycrystalline material \cite{Ribeiro}. The average value of the D/G intensity ratio is 0.32~$\pm$~0.10, 0.27~$\pm$~0.10, and 0.26~$\pm$~0.10 for the different areas. These values correspond to an average grain size of graphene ranging from 70~nm to 100~nm \cite{Ribeiro}.

In order to evaluate the strain-doping effects in the graphene film, Figure \ref{raman1}(e) presents the correlation plot of the Raman shift of the G peak with the Raman shift of the 2D mode. The dispersion of the different areas confirms a homogeneous n-type doping of the graphene, while there is an important spreading of the data along the strain direction. The slope of the linear fits for the different dispersions from areas 1 to 3 varies between 1.77 and 1.90. This is a clear indication for biaxial strain in the graphene layer, highly expected due to the complex nature of the substrate morphology.

In Figure \ref{raman1}(a), we report the absence of the second order SiC-related Raman modes. In order to clarify the absence of these peaks, we analyze the range of Raman shifts related to the transverse optical (TO) phonons of 4H-SiC, reported in Figure S4. These peaks are well known to be related to the crystallinity of the SiC substrate. On the pristine Si-face 4H-SiC wafer, the Raman analysis shows prominent TO peaks and the presence of second order peaks, as reported in Figure S4. Figures S4 and S5 show that the porousification process and the high temperature process have a strong impact on the 4H-SiC substrate, as witnessed by a strong suppression of the intensity of the TO Raman modes at 775~cm$^{-1}$ and 796~cm$^{-1}$, respectively. However, the peak widths demonstrate that the crystal quality of the porousified material is not deeply affected, and the peak widths and positions are in agreement with bulk crystals. The statistical analysis of the 4H-SiC TO peak is reported in Figure S7.

\textbf{C-face samples.} We have also investigated the graphene growth on the C-face of 4H-SiC. Several samples have been annealed at different temperatures, in order to find the best growth conditions. With respect to the Si-face, the C-face demonstrates a much more critical dependence on temperature, producing high quality graphene in a narrow range of about $40^\circ$~C. The best graphene growth on the C-face samples has been achieved by annealing at $1260^\circ$~C, which is lower than the optimal growth temperature on the Si-face. 

%\subsection{C-face bulk analysis}

\begin{figure*}
   \begin{center}
   \includegraphics[width=0.9\textwidth]{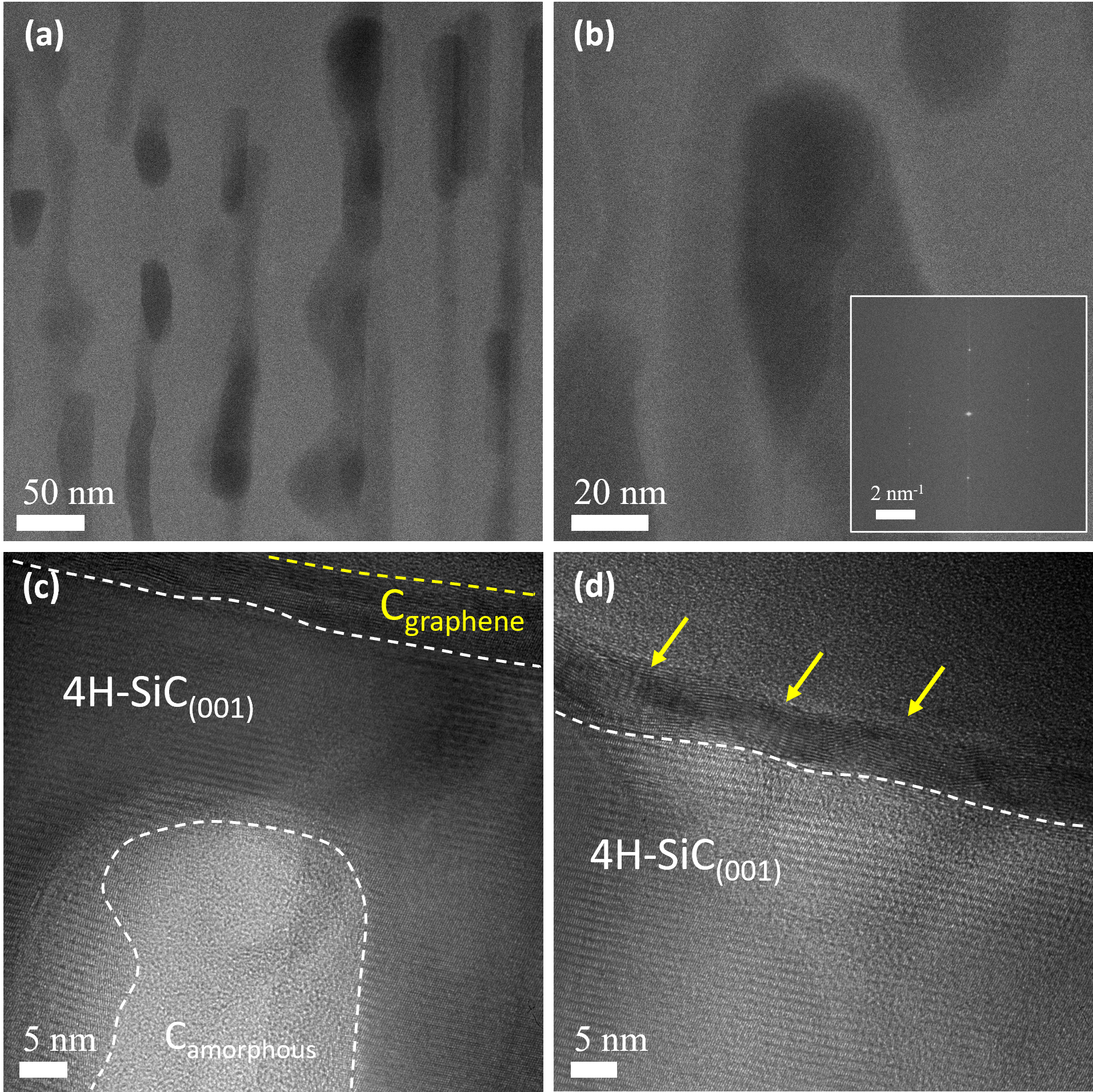}
   \end{center}
\caption{\label{TEM2} High resolution TEM images of a C-face sample after graphene growth at $1260^\circ$~C. (a) TEM image at 80 kV on an area 395~nm $\times$ 395~nm showing multiple pores, (b) magnification at 140~nm $\times$ 140~nm on one of the pores, (c) magnification at 60~nm $\times$ 60~nm on a single pore, filled with amorphous carbon, (d) magnification at 60~nm $\times$ 60~nm. This last image shows the presence of graphene sheets on the sample surface only, marked by yellow arrows.}
\end{figure*}

As reported in Figure \ref{SEM}, the pore structure in C-face samples is quite different from that in Si-face samples. For the C-face, pores are distributed along "channels" starting from the surface toward the inner of the sample. Moreover, pores  are rarely interconnected from channel to channel. At larger magnification, the channel structure is evident as shown in Figure \ref{TEM2}(a). Figure \ref{TEM2}(b), taken at a higher magnification, shows the elongated structure of a single pore. The FFT of this image does not show any extra spots (see inset on Figure \ref{TEM2}(b)), in contrast to the Si-face sample. Moreover, pores are rarely found empty, but filled with amorphous carbon as shown in Figure \ref{TEM2}(c). We attribute this fact to a plug effect. The pores are plugged by the fast Si evaporation rate, which results in a carbon accumulation in the pores, preventing the formation of graphene. Indeed, in C-face samples, no graphene sheets were observed inside the pores (see Figure \ref{TEM2}(c)), while a graphene sheet stacking on the sample surface is observed, as shown in Figure \ref{TEM2}(d), marked by yellow arrows.

\begin{figure*}
   \begin{center}
   \includegraphics[width=0.9\textwidth]{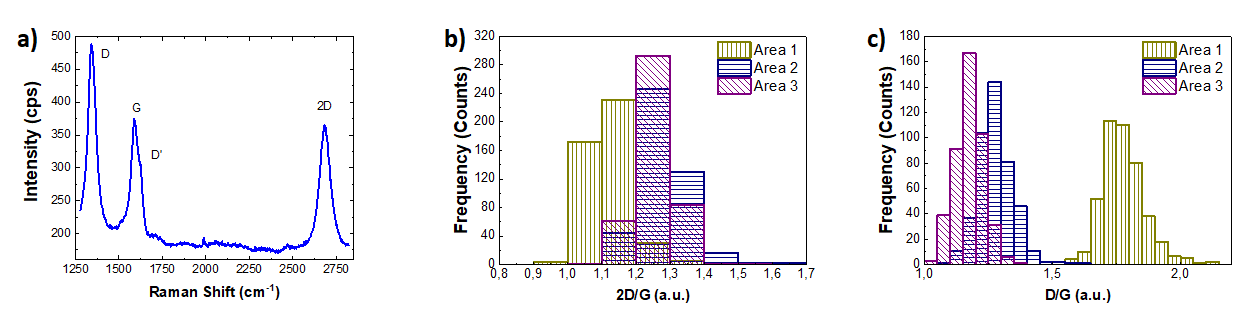}
   \end{center}
\caption{\label{raman2} Raman analysis of a porous 4H-SiC C-face sample after annealing at $1260^\circ$~C (a) Raman spectrum from 1250 to 2850~cm$^{-1}$ (graphene peaks), (b) superposition of the 2D/G ratio histograms of the areas 1, 2 and 3, (c) superposition of the D/G ratio histograms of the areas 1, 2 and 3.}
\end{figure*}

Raman analysis of these graphene layers is reported in Figure \ref{raman2}. At a glance, it is evident that the graphene is more defective and less homogeneous than for the Si-face samples. Indeed, in a representative Raman spectrum shown in Figure \ref{raman2}(a), the intensities of the D and D’ Raman modes, peaked at 1341~cm$^{-1}$ and 1615~cm$^{-1}$, are more intense, demonstrating a more defective graphene. The 2D mode is peaked at 2681~cm$^{-1}$ with a FWHM of 75~cm$^{-1}$. The G peak appears at 1588~cm$^{-1}$. Raman mapping is carried out in three different areas, as indicated in Figures S12 - S17. The statistical analysis of the Raman maps, evaluating the different graphene benchmark parameters, is reported in Figures \ref{raman2}(b) and (c).  In particular, Figure \ref{raman2}(b) presents histograms of the 2D/G intensity ratio, obtained in the different areas. The average values are 1.12~$\pm$~0.1, 1.27~$\pm$~0.1 and 1.25~$\pm$~0.1 for area 1, 2 and 3, respectively. These values suggest bilayer formation \cite{Lee}. This claim is supported by the 2D FWHM analysis, reported in Figure S14. The D/G intensity ratio demonstrates a different grain size of graphene in area 1 with respect to areas 2 and 3. In fact, the average values of the D/G intensity ratio are 1.76~$\pm$~0.1 (area 1), 1.27~$\pm$~0.05 (area 2) and 1.18~$\pm$~0.05 (area 3), yielding an average grain size of 2~nm in area 1, while in the other two areas the grain size is between 5~nm and 7~nm, about an order of magnitude smaller than for the Si-face. As will be discussed later, the observed inhomogeneities in these C-face samples are due to a temperature gradient in the sample that is large enough to produce an observable variation.

As reported for the Si-face, before the high temperature process, the Raman analysis of the porousified C-face 4H-SiC substrate shows prominent TO and LO peaks and the presence of second order peaks, as reported in Figure S11 of the Supporting Information. The Raman spectrum in the SiC range for the sample after graphene growth is reported in Figure S11(a). Even in this case, the 4H-SiC TO and LO modes are strongly suppressed in intensity. The statistical analysis of the TO peak is reported in Figure S11(b). In addition, broad bands appear at about 300~cm$^{-1}$ and 800~cm$^{-1}$ (see S13(a)), indicating an amorphization of the SiC substrate \cite{inoue}. 

 The growth temperatures reported here are in agreement with those reported for 2D growth of graphene in 4H-SiC, between 1100 and $1450^\circ$~C, with annealing times in the minutes range \cite{Forbeaux, Hass, Berger, Ouerghi}. In 2D samples it was found that the graphene quality increases if the Si surface is encapsulated, leaving a small calibrated leak which slows down the Si evaporation rate. This technique is referred to as confinement controlled sublimation (CCS)\cite{deHeer}. The porous SiC samples benefit from the CCS effect, since the pores confine the Si and act as leak regulators. Inside the pores, the graphene growth is also affected by the different crystal planes of the pore walls. Therefore, it not surprising that the graphene is stacked inside the pores, as observed in the HAADF-STEM images on Figure \ref{TEM}. 

The  growth of 2D graphene on C-face 4H-SiC is faster and less controllable than on the Si-face, and usually thick multilayers grow \cite{Giusca}. Moreover layers are made from rotationally disordered graphene sheets, producing a turbostratic graphene. Therefore, the 2D growth of a monolayer graphene from a C-face substrate is still a challenge.  The lower growth temperature on the C-face with respect to the Si-face found in this work is in agreement with literature data \cite{Luxmi}. 

On C-face 4H-SiC, the graphene growth process can still be optimized. Indeed, the graphene growth is less homogeneous at the surface, and most pores are filled with amorphous carbon. This is due to several factors. The pores of the C-face samples are smaller, and the growth window is narrower. We expect that a modified porosification protocol \cite{Leitgeb2021} for the C-face samples will result in larger pores, reducing the plugging effect. Moreover, an evaluation of the sample temperature along the length of the sample demonstrated a gradient of about $50^\circ$~C (see supporting information for details). Such a gradient does not impact the growth in Si-face samples, which feature a larger growth window, but it strongly impacts the graphene growth on C-face samples, which have a narrower growth window. However, it will be straightforward to reduce the temperature gradient on the sample, which will result in a higher homogeneity of the graphene also on C-face samples.

\section{Conclusions}
%\label{}
We have demonstrated the growth of high quality epitaxial graphene on a nano-porous crystalline 4H-SiC(0001) for the first time. The porosification process and the graphene growth has been performed in both Si-face and C-face orientation. Porous samples were analyzed by SEM, TEM, electron tomography and Raman spectroscopy, in order to characterize the surface and pores structure and the graphene quality. The electron tomography allowed to measure an average pore diameter of 180~nm and 85~nm for Si-face and C-face samples respectively. The TEM analysis has demonstrated the graphene growth inside pores of Si-face samples and the Raman spectroscopy confirmed the high quality of graphene, with a 2D/G ratio around 1 for both Si-face and C-face samples. The tunability of pore size and the scalability of the porosification process to large surface samples increase the flexibility of this material. The high surface-to-volume ratio and the native three-dimensional graphene on a semiconductor backbone represent a breakthrough for a huge number of applications in many fields as nano-electronic, sensors, energy materials and membrane technology. 

\begin{acknowledgments}
The authors want to acknowledge L. Covaci for helpful discussion. The authors acknowledge funding from the EU (ERC Consolidator Grant REALNANO 815128 and Grant Agreement No. 731019 (EUSMI) under the Horizon 2020 Programme).
\end{acknowledgments}

%\nocite{*}
\bibliography{biblio}% Produces the bibliography via BibTeX.

%aipnum4-2.bst 2019-01-14 (MD) hand-edited version of apsrev4-1.bst
\begin{thebibliography}{62}%
\makeatletter
\providecommand \@ifxundefined [1]{%
 \@ifx{#1\undefined}
}%
\providecommand \@ifnum [1]{%
 \ifnum #1\expandafter \@firstoftwo
 \else \expandafter \@secondoftwo
 \fi
}%
\providecommand \@ifx [1]{%
 \ifx #1\expandafter \@firstoftwo
 \else \expandafter \@secondoftwo
 \fi
}%
\providecommand \natexlab [1]{#1}%
\providecommand \enquote  [1]{``#1''}%
\providecommand \bibnamefont  [1]{#1}%
\providecommand \bibfnamefont [1]{#1}%
\providecommand \citenamefont [1]{#1}%
\providecommand \href@noop [0]{\@secondoftwo}%
\providecommand \href [0]{\begingroup \@sanitize@url \@href}%
\providecommand \@href[1]{\@@startlink{#1}\@@href}%
\providecommand \@@href[1]{\endgroup#1\@@endlink}%
\providecommand \@sanitize@url [0]{\catcode `\\12\catcode `\$12\catcode
  `\&12\catcode `\#12\catcode `\^12\catcode `\_12\catcode `\%12\relax}%
\providecommand \@@startlink[1]{}%
\providecommand \@@endlink[0]{}%
\providecommand \url  [0]{\begingroup\@sanitize@url \@url }%
\providecommand \@url [1]{\endgroup\@href {#1}{\urlprefix }}%
\providecommand \urlprefix  [0]{URL }%
\providecommand \Eprint [0]{\href }%
\providecommand \doibase [0]{https://doi.org/}%
\providecommand \selectlanguage [0]{\@gobble}%
\providecommand \bibinfo  [0]{\@secondoftwo}%
\providecommand \bibfield  [0]{\@secondoftwo}%
\providecommand \translation [1]{[#1]}%
\providecommand \BibitemOpen [0]{}%
\providecommand \bibitemStop [0]{}%
\providecommand \bibitemNoStop [0]{.\EOS\space}%
\providecommand \EOS [0]{\spacefactor3000\relax}%
\providecommand \BibitemShut  [1]{\csname bibitem#1\endcsname}%
\let\auto@bib@innerbib\@empty
%</preamble>
\bibitem [{\citenamefont {Wang}\ \emph {et~al.}(2019)\citenamefont {Wang},
  \citenamefont {Mu}, \citenamefont {Sun},\ and\ \citenamefont {Mu}}]{Wang}%
  \BibitemOpen
  \bibfield  {author} {\bibinfo {author} {\bibfnamefont {J.}~\bibnamefont
  {Wang}}, \bibinfo {author} {\bibfnamefont {X.}~\bibnamefont {Mu}}, \bibinfo
  {author} {\bibfnamefont {M.}~\bibnamefont {Sun}},\ and\ \bibinfo {author}
  {\bibfnamefont {T.}~\bibnamefont {Mu}},\ }\bibfield  {title} {\enquote
  {\bibinfo {title} {Optoelectronic properties and applications of
  graphene-based hybrid nanomaterials and van der waals heterostructures},}\
  }\href {https://doi.org/10.1016/j.apmt.2019.03.006} {\bibfield  {journal}
  {\bibinfo  {journal} {Applied Materials Today}\ }\textbf {\bibinfo {volume}
  {16}},\ \bibinfo {pages} {1--20} (\bibinfo {year} {2019})}\BibitemShut
  {NoStop}%
\bibitem [{\citenamefont {Shrivas}\ \emph {et~al.}(2020)\citenamefont
  {Shrivas}, \citenamefont {Ghosale}, \citenamefont {Bajpai}, \citenamefont
  {Kant}, \citenamefont {Dewangan},\ and\ \citenamefont {Shankar}}]{Shrivas}%
  \BibitemOpen
  \bibfield  {author} {\bibinfo {author} {\bibfnamefont {K.}~\bibnamefont
  {Shrivas}}, \bibinfo {author} {\bibfnamefont {A.}~\bibnamefont {Ghosale}},
  \bibinfo {author} {\bibfnamefont {P.~K.}\ \bibnamefont {Bajpai}}, \bibinfo
  {author} {\bibfnamefont {T.}~\bibnamefont {Kant}}, \bibinfo {author}
  {\bibfnamefont {K.}~\bibnamefont {Dewangan}},\ and\ \bibinfo {author}
  {\bibfnamefont {R.}~\bibnamefont {Shankar}},\ }\bibfield  {title} {\enquote
  {\bibinfo {title} {Advances in flexible electronics and electrochemical
  sensors using conducting nanomaterials: A review},}\ }\href
  {https://doi.org/10.1016/j.microc.2020.104944} {\bibfield  {journal}
  {\bibinfo  {journal} {Microchemical Journal}\ }\textbf {\bibinfo {volume}
  {156}},\ \bibinfo {pages} {104944} (\bibinfo {year} {2020})}\BibitemShut
  {NoStop}%
\bibitem [{\citenamefont {Singh}, \citenamefont {Meyyappan},\ and\
  \citenamefont {Nalwa}(2017)}]{Singh}%
  \BibitemOpen
  \bibfield  {author} {\bibinfo {author} {\bibfnamefont {E.}~\bibnamefont
  {Singh}}, \bibinfo {author} {\bibfnamefont {M.}~\bibnamefont {Meyyappan}},\
  and\ \bibinfo {author} {\bibfnamefont {H.~S.}\ \bibnamefont {Nalwa}},\
  }\bibfield  {title} {\enquote {\bibinfo {title} {{Flexible Graphene-Based
  Wearable Gas and Chemical Sensors}},}\ }\href
  {https://doi.org/10.1021/acsami.7b07063} {\bibfield  {journal} {\bibinfo
  {journal} {ACS Applied Materials \& Interfaces}\ }\textbf {\bibinfo {volume}
  {9}},\ \bibinfo {pages} {34544--34586} (\bibinfo {year} {2017})}\BibitemShut
  {NoStop}%
\bibitem [{\citenamefont {Nag}, \citenamefont {Mitra},\ and\ \citenamefont
  {Mukhopadhyay}(2018)}]{Nag}%
  \BibitemOpen
  \bibfield  {author} {\bibinfo {author} {\bibfnamefont {A.}~\bibnamefont
  {Nag}}, \bibinfo {author} {\bibfnamefont {A.}~\bibnamefont {Mitra}},\ and\
  \bibinfo {author} {\bibfnamefont {S.}~\bibnamefont {Mukhopadhyay}},\
  }\bibfield  {title} {\enquote {\bibinfo {title} {Graphene and its
  sensor-based applications: A review},}\ }\href
  {https://doi.org/https://doi.org/10.1016/j.sna.2017.12.028} {\bibfield
  {journal} {\bibinfo  {journal} {Sensors and Actuators A: Physical}\ }\textbf
  {\bibinfo {volume} {270}},\ \bibinfo {pages} {177--194} (\bibinfo {year}
  {2018})}\BibitemShut {NoStop}%
\bibitem [{\citenamefont {Chauhan}, \citenamefont {Maekawa},\ and\
  \citenamefont {Kumar}(2017)}]{Kumar}%
  \BibitemOpen
  \bibfield  {author} {\bibinfo {author} {\bibfnamefont {N.}~\bibnamefont
  {Chauhan}}, \bibinfo {author} {\bibfnamefont {T.}~\bibnamefont {Maekawa}},\
  and\ \bibinfo {author} {\bibfnamefont {D.~N.~S.}\ \bibnamefont {Kumar}},\
  }\bibfield  {title} {\enquote {\bibinfo {title} {Graphene based
  biosensors—accelerating medical diagnostics to new-dimensions},}\ }\href
  {https://doi.org/10.1557/jmr.2017.91} {\bibfield  {journal} {\bibinfo
  {journal} {Journal of Materials Research}\ }\textbf {\bibinfo {volume}
  {32}},\ \bibinfo {pages} {2860–2882} (\bibinfo {year} {2017})}\BibitemShut
  {NoStop}%
\bibitem [{\citenamefont {Wang}\ \emph {et~al.}(2020)\citenamefont {Wang},
  \citenamefont {Sun}, \citenamefont {Su}, \citenamefont {Wang},\ and\
  \citenamefont {Lv}}]{WangC}%
  \BibitemOpen
  \bibfield  {author} {\bibinfo {author} {\bibfnamefont {C.-F.}\ \bibnamefont
  {Wang}}, \bibinfo {author} {\bibfnamefont {X.-Y.}\ \bibnamefont {Sun}},
  \bibinfo {author} {\bibfnamefont {M.}~\bibnamefont {Su}}, \bibinfo {author}
  {\bibfnamefont {Y.-P.}\ \bibnamefont {Wang}},\ and\ \bibinfo {author}
  {\bibfnamefont {Y.-K.}\ \bibnamefont {Lv}},\ }\bibfield  {title} {\enquote
  {\bibinfo {title} {Electrochemical biosensors based on antibody{,} nucleic
  acid and enzyme functionalized graphene for the detection of disease-related
  biomolecules},}\ }\href {https://doi.org/10.1039/C9AN02047K} {\bibfield
  {journal} {\bibinfo  {journal} {Analyst}\ }\textbf {\bibinfo {volume}
  {145}},\ \bibinfo {pages} {1550--1562} (\bibinfo {year} {2020})}\BibitemShut
  {NoStop}%
\bibitem [{\citenamefont {Merko{\c{c}}i}(2020)}]{Merkoci}%
  \BibitemOpen
  \bibfield  {author} {\bibinfo {author} {\bibfnamefont {A.}~\bibnamefont
  {Merko{\c{c}}i}},\ }\bibfield  {title} {\enquote {\bibinfo {title}
  {Graphene-based biosensors},}\ }\href
  {https://doi.org/10.1088/2053-1583/aba3bf} {\bibfield  {journal} {\bibinfo
  {journal} {2D Materials}\ }\textbf {\bibinfo {volume} {7}},\ \bibinfo {pages}
  {040401} (\bibinfo {year} {2020})}\BibitemShut {NoStop}%
\bibitem [{\citenamefont {Boukhvalov}\ and\ \citenamefont
  {Katsnelson}(2008)}]{Bouk}%
  \BibitemOpen
  \bibfield  {author} {\bibinfo {author} {\bibfnamefont {D.~W.}\ \bibnamefont
  {Boukhvalov}}\ and\ \bibinfo {author} {\bibfnamefont {M.~I.}\ \bibnamefont
  {Katsnelson}},\ }\bibfield  {title} {\enquote {\bibinfo {title} {Chemical
  functionalization of graphene with defects},}\ }\href
  {https://doi.org/10.1021/nl802234n} {\bibfield  {journal} {\bibinfo
  {journal} {Nano Letters}\ }\textbf {\bibinfo {volume} {8}},\ \bibinfo {pages}
  {4373--4379} (\bibinfo {year} {2008})}\BibitemShut {NoStop}%
\bibitem [{\citenamefont {Ortiz-Medina}\ \emph {et~al.}(2019)\citenamefont
  {Ortiz-Medina}, \citenamefont {Wang}, \citenamefont {Cruz-Silva},
  \citenamefont {Morelos-Gomez}, \citenamefont {Wang}, \citenamefont {Yao},
  \citenamefont {Terrones},\ and\ \citenamefont {Endo}}]{Ortiz}%
  \BibitemOpen
  \bibfield  {author} {\bibinfo {author} {\bibfnamefont {J.}~\bibnamefont
  {Ortiz-Medina}}, \bibinfo {author} {\bibfnamefont {Z.}~\bibnamefont {Wang}},
  \bibinfo {author} {\bibfnamefont {R.}~\bibnamefont {Cruz-Silva}}, \bibinfo
  {author} {\bibfnamefont {A.}~\bibnamefont {Morelos-Gomez}}, \bibinfo {author}
  {\bibfnamefont {F.}~\bibnamefont {Wang}}, \bibinfo {author} {\bibfnamefont
  {X.}~\bibnamefont {Yao}}, \bibinfo {author} {\bibfnamefont {M.}~\bibnamefont
  {Terrones}},\ and\ \bibinfo {author} {\bibfnamefont {M.}~\bibnamefont
  {Endo}},\ }\bibfield  {title} {\enquote {\bibinfo {title} {{Defect
  Engineering and Surface Functionalization of Nanocarbons for Metal-Free
  Catalysis}},}\ }\href
  {https://doi.org/https://doi.org/10.1002/adma.201805717} {\bibfield
  {journal} {\bibinfo  {journal} {Advanced Materials}\ }\textbf {\bibinfo
  {volume} {31}},\ \bibinfo {pages} {1805717} (\bibinfo {year}
  {2019})}\BibitemShut {NoStop}%
\bibitem [{\citenamefont {Ye}\ \emph {et~al.}(2017)\citenamefont {Ye},
  \citenamefont {Cai}, \citenamefont {Yang}, \citenamefont {Tang},
  \citenamefont {Zhou}, \citenamefont {Tan}, \citenamefont {Xie},\ and\
  \citenamefont {Zheng}}]{Ye}%
  \BibitemOpen
  \bibfield  {author} {\bibinfo {author} {\bibfnamefont {X.-L.}\ \bibnamefont
  {Ye}}, \bibinfo {author} {\bibfnamefont {J.}~\bibnamefont {Cai}}, \bibinfo
  {author} {\bibfnamefont {X.-D.}\ \bibnamefont {Yang}}, \bibinfo {author}
  {\bibfnamefont {X.-Y.}\ \bibnamefont {Tang}}, \bibinfo {author}
  {\bibfnamefont {Z.-Y.}\ \bibnamefont {Zhou}}, \bibinfo {author}
  {\bibfnamefont {Y.-Z.}\ \bibnamefont {Tan}}, \bibinfo {author} {\bibfnamefont
  {S.-Y.}\ \bibnamefont {Xie}},\ and\ \bibinfo {author} {\bibfnamefont {L.-S.}\
  \bibnamefont {Zheng}},\ }\bibfield  {title} {\enquote {\bibinfo {title}
  {Quantifying defect-enhanced chemical functionalization of single-layer
  graphene and its application in supramolecular assembly},}\ }\href
  {https://doi.org/10.1039/C7TA07612F} {\bibfield  {journal} {\bibinfo
  {journal} {J. Mater. Chem. A}\ }\textbf {\bibinfo {volume} {5}},\ \bibinfo
  {pages} {24257--24262} (\bibinfo {year} {2017})}\BibitemShut {NoStop}%
\bibitem [{\citenamefont {Yan}\ \emph {et~al.}(2020)\citenamefont {Yan},
  \citenamefont {Shin}, \citenamefont {Chen}, \citenamefont {Lee},
  \citenamefont {Manickam}, \citenamefont {Hanson}, \citenamefont {Zhao},
  \citenamefont {Lester}, \citenamefont {Wu},\ and\ \citenamefont
  {Pang}}]{Yan2020}%
  \BibitemOpen
  \bibfield  {author} {\bibinfo {author} {\bibfnamefont {Y.}~\bibnamefont
  {Yan}}, \bibinfo {author} {\bibfnamefont {W.~I.}\ \bibnamefont {Shin}},
  \bibinfo {author} {\bibfnamefont {H.}~\bibnamefont {Chen}}, \bibinfo {author}
  {\bibfnamefont {S.-M.}\ \bibnamefont {Lee}}, \bibinfo {author} {\bibfnamefont
  {S.}~\bibnamefont {Manickam}}, \bibinfo {author} {\bibfnamefont
  {S.}~\bibnamefont {Hanson}}, \bibinfo {author} {\bibfnamefont
  {H.}~\bibnamefont {Zhao}}, \bibinfo {author} {\bibfnamefont {E.}~\bibnamefont
  {Lester}}, \bibinfo {author} {\bibfnamefont {T.}~\bibnamefont {Wu}},\ and\
  \bibinfo {author} {\bibfnamefont {C.~H.}\ \bibnamefont {Pang}},\ }\bibfield
  {title} {\enquote {\bibinfo {title} {A recent trend: application of graphene
  in catalysis},}\ }\href {https://doi.org/10.1007/s42823-020-00200-7}
  {\bibfield  {journal} {\bibinfo  {journal} {Carbon Letters}\ }\textbf
  {\bibinfo {volume} {31}},\ \bibinfo {pages} {177--199} (\bibinfo {year}
  {2020})}\BibitemShut {NoStop}%
\bibitem [{\citenamefont {Sun}\ \emph {et~al.}(2015)\citenamefont {Sun},
  \citenamefont {Liu}, \citenamefont {Liu}, \citenamefont {Qu},\ and\
  \citenamefont {Li}}]{Sun}%
  \BibitemOpen
  \bibfield  {author} {\bibinfo {author} {\bibfnamefont {M.}~\bibnamefont
  {Sun}}, \bibinfo {author} {\bibfnamefont {H.}~\bibnamefont {Liu}}, \bibinfo
  {author} {\bibfnamefont {Y.}~\bibnamefont {Liu}}, \bibinfo {author}
  {\bibfnamefont {J.}~\bibnamefont {Qu}},\ and\ \bibinfo {author}
  {\bibfnamefont {J.}~\bibnamefont {Li}},\ }\bibfield  {title} {\enquote
  {\bibinfo {title} {Graphene-based transition metal oxide nanocomposites for
  the oxygen reduction reaction},}\ }\href {https://doi.org/10.1039/C4NR05838K}
  {\bibfield  {journal} {\bibinfo  {journal} {Nanoscale}\ }\textbf {\bibinfo
  {volume} {7}},\ \bibinfo {pages} {1250--1269} (\bibinfo {year}
  {2015})}\BibitemShut {NoStop}%
\bibitem [{\citenamefont {Wang}\ \emph {et~al.}(2015)\citenamefont {Wang},
  \citenamefont {Huang}, \citenamefont {Yang}, \citenamefont {Xu},
  \citenamefont {He}, \citenamefont {Li}, \citenamefont {Hu}, \citenamefont
  {Yin}, \citenamefont {He},\ and\ \citenamefont {Zhang}}]{wang2015}%
  \BibitemOpen
  \bibfield  {author} {\bibinfo {author} {\bibfnamefont {T.}~\bibnamefont
  {Wang}}, \bibinfo {author} {\bibfnamefont {D.}~\bibnamefont {Huang}},
  \bibinfo {author} {\bibfnamefont {Z.}~\bibnamefont {Yang}}, \bibinfo {author}
  {\bibfnamefont {S.}~\bibnamefont {Xu}}, \bibinfo {author} {\bibfnamefont
  {G.}~\bibnamefont {He}}, \bibinfo {author} {\bibfnamefont {X.}~\bibnamefont
  {Li}}, \bibinfo {author} {\bibfnamefont {N.}~\bibnamefont {Hu}}, \bibinfo
  {author} {\bibfnamefont {G.}~\bibnamefont {Yin}}, \bibinfo {author}
  {\bibfnamefont {D.}~\bibnamefont {He}},\ and\ \bibinfo {author}
  {\bibfnamefont {L.}~\bibnamefont {Zhang}},\ }\bibfield  {title} {\enquote
  {\bibinfo {title} {A review on graphene-based gas/vapor sensors with unique
  properties and potential applications},}\ }\href
  {https://doi.org/10.1007/s40820-015-0073-1} {\bibfield  {journal} {\bibinfo
  {journal} {Nano-Micro Letters}\ }\textbf {\bibinfo {volume} {8}},\ \bibinfo
  {pages} {95--119} (\bibinfo {year} {2015})}\BibitemShut {NoStop}%
\bibitem [{\citenamefont {Mohan}\ \emph {et~al.}(2019)\citenamefont {Mohan},
  \citenamefont {Sharma}, \citenamefont {Kumar},\ and\ \citenamefont
  {Gayathri}}]{Mohan2019}%
  \BibitemOpen
  \bibfield  {author} {\bibinfo {author} {\bibfnamefont {M.}~\bibnamefont
  {Mohan}}, \bibinfo {author} {\bibfnamefont {V.~K.}\ \bibnamefont {Sharma}},
  \bibinfo {author} {\bibfnamefont {E.~A.}\ \bibnamefont {Kumar}},\ and\
  \bibinfo {author} {\bibfnamefont {V.}~\bibnamefont {Gayathri}},\ }\bibfield
  {title} {\enquote {\bibinfo {title} {Hydrogen storage in carbon
  materials{\textemdash}a review},}\ }\href {https://doi.org/10.1002/est2.35}
  {\bibfield  {journal} {\bibinfo  {journal} {Energy Storage}\ }\textbf
  {\bibinfo {volume} {1}},\ \bibinfo {pages} {e35} (\bibinfo {year}
  {2019})}\BibitemShut {NoStop}%
\bibitem [{\citenamefont {Jain}\ and\ \citenamefont
  {Kandasubramanian}(2019)}]{Jain2019}%
  \BibitemOpen
  \bibfield  {author} {\bibinfo {author} {\bibfnamefont {V.}~\bibnamefont
  {Jain}}\ and\ \bibinfo {author} {\bibfnamefont {B.}~\bibnamefont
  {Kandasubramanian}},\ }\bibfield  {title} {\enquote {\bibinfo {title}
  {Functionalized graphene materials for hydrogen storage},}\ }\href
  {https://doi.org/10.1007/s10853-019-04150-y} {\bibfield  {journal} {\bibinfo
  {journal} {Journal of Materials Science}\ }\textbf {\bibinfo {volume} {55}},\
  \bibinfo {pages} {1865--1903} (\bibinfo {year} {2019})}\BibitemShut {NoStop}%
\bibitem [{\citenamefont {McCallion}\ \emph {et~al.}(2016)\citenamefont
  {McCallion}, \citenamefont {Burthem}, \citenamefont {Rees-Unwin},
  \citenamefont {Golovanov},\ and\ \citenamefont {Pluen}}]{McCallion}%
  \BibitemOpen
  \bibfield  {author} {\bibinfo {author} {\bibfnamefont {C.}~\bibnamefont
  {McCallion}}, \bibinfo {author} {\bibfnamefont {J.}~\bibnamefont {Burthem}},
  \bibinfo {author} {\bibfnamefont {K.}~\bibnamefont {Rees-Unwin}}, \bibinfo
  {author} {\bibfnamefont {A.}~\bibnamefont {Golovanov}},\ and\ \bibinfo
  {author} {\bibfnamefont {A.}~\bibnamefont {Pluen}},\ }\bibfield  {title}
  {\enquote {\bibinfo {title} {Graphene in therapeutics delivery: Problems,
  solutions and future opportunities},}\ }\href
  {https://doi.org/https://doi.org/10.1016/j.ejpb.2016.04.015} {\bibfield
  {journal} {\bibinfo  {journal} {European Journal of Pharmaceutics and
  Biopharmaceutics}\ }\textbf {\bibinfo {volume} {104}},\ \bibinfo {pages}
  {235--250} (\bibinfo {year} {2016})}\BibitemShut {NoStop}%
\bibitem [{\citenamefont {Tiliakos}\ \emph {et~al.}(2020)\citenamefont
  {Tiliakos}, \citenamefont {Trefilov}, \citenamefont {Tanasă}, \citenamefont
  {Balan},\ and\ \citenamefont {Stamatin}}]{Tiliakos}%
  \BibitemOpen
  \bibfield  {author} {\bibinfo {author} {\bibfnamefont {A.}~\bibnamefont
  {Tiliakos}}, \bibinfo {author} {\bibfnamefont {A.~M.}\ \bibnamefont
  {Trefilov}}, \bibinfo {author} {\bibfnamefont {E.}~\bibnamefont {Tanasă}},
  \bibinfo {author} {\bibfnamefont {A.}~\bibnamefont {Balan}},\ and\ \bibinfo
  {author} {\bibfnamefont {I.}~\bibnamefont {Stamatin}},\ }\bibfield  {title}
  {\enquote {\bibinfo {title} {Laser-induced graphene as the microporous layer
  in proton exchange membrane fuel cells},}\ }\href
  {https://doi.org/https://doi.org/10.1016/j.apsusc.2019.144096} {\bibfield
  {journal} {\bibinfo  {journal} {Applied Surface Science}\ }\textbf {\bibinfo
  {volume} {504}},\ \bibinfo {pages} {144096} (\bibinfo {year}
  {2020})}\BibitemShut {NoStop}%
\bibitem [{\citenamefont {Thimmappa}\ \emph {et~al.}(2019)\citenamefont
  {Thimmappa}, \citenamefont {Gautam}, \citenamefont {Devendrachari},
  \citenamefont {Kottaichamy}, \citenamefont {Bhat}, \citenamefont {Umar},\
  and\ \citenamefont {Thotiyl}}]{Thimmappa}%
  \BibitemOpen
  \bibfield  {author} {\bibinfo {author} {\bibfnamefont {R.}~\bibnamefont
  {Thimmappa}}, \bibinfo {author} {\bibfnamefont {M.}~\bibnamefont {Gautam}},
  \bibinfo {author} {\bibfnamefont {M.~C.}\ \bibnamefont {Devendrachari}},
  \bibinfo {author} {\bibfnamefont {A.~R.}\ \bibnamefont {Kottaichamy}},
  \bibinfo {author} {\bibfnamefont {Z.~M.}\ \bibnamefont {Bhat}}, \bibinfo
  {author} {\bibfnamefont {A.}~\bibnamefont {Umar}},\ and\ \bibinfo {author}
  {\bibfnamefont {M.~O.}\ \bibnamefont {Thotiyl}},\ }\bibfield  {title}
  {\enquote {\bibinfo {title} {Proton-conducting graphene membrane electrode
  assembly for high performance hydrogen fuel cells},}\ }\href
  {https://doi.org/10.1021/acssuschemeng.9b02917} {\bibfield  {journal}
  {\bibinfo  {journal} {ACS Sustainable Chemistry \& Engineering}\ }\textbf
  {\bibinfo {volume} {7}},\ \bibinfo {pages} {14189--14194} (\bibinfo {year}
  {2019})}\BibitemShut {NoStop}%
\bibitem [{\citenamefont {Venkateshalu}\ and\ \citenamefont
  {Grace}(2020)}]{Venkateshalu}%
  \BibitemOpen
  \bibfield  {author} {\bibinfo {author} {\bibfnamefont {S.}~\bibnamefont
  {Venkateshalu}}\ and\ \bibinfo {author} {\bibfnamefont {A.~N.}\ \bibnamefont
  {Grace}},\ }\bibfield  {title} {\enquote {\bibinfo {title}
  {Review{\textemdash}heterogeneous 3d graphene derivatives for
  supercapacitors},}\ }\href {https://doi.org/10.1149/1945-7111/ab6bc5}
  {\bibfield  {journal} {\bibinfo  {journal} {Journal of The Electrochemical
  Society}\ }\textbf {\bibinfo {volume} {167}},\ \bibinfo {pages} {050509}
  (\bibinfo {year} {2020})}\BibitemShut {NoStop}%
\bibitem [{\citenamefont {Ping}\ \emph {et~al.}(2017)\citenamefont {Ping},
  \citenamefont {Gong}, \citenamefont {Fu},\ and\ \citenamefont
  {Pan}}]{Ping17}%
  \BibitemOpen
  \bibfield  {author} {\bibinfo {author} {\bibfnamefont {Y.}~\bibnamefont
  {Ping}}, \bibinfo {author} {\bibfnamefont {Y.}~\bibnamefont {Gong}}, \bibinfo
  {author} {\bibfnamefont {Q.}~\bibnamefont {Fu}},\ and\ \bibinfo {author}
  {\bibfnamefont {C.}~\bibnamefont {Pan}},\ }\bibfield  {title} {\enquote
  {\bibinfo {title} {Preparation of three-dimensional graphene foam for high
  performance supercapacitors},}\ }\href
  {https://doi.org/https://doi.org/10.1016/j.pnsc.2017.03.005} {\bibfield
  {journal} {\bibinfo  {journal} {Progress in Natural Science: Materials
  International}\ }\textbf {\bibinfo {volume} {27}},\ \bibinfo {pages}
  {177--181} (\bibinfo {year} {2017})}\BibitemShut {NoStop}%
\bibitem [{\citenamefont {Shen}\ \emph {et~al.}(2021)\citenamefont {Shen},
  \citenamefont {Sun}, \citenamefont {Yang}, \citenamefont {Krasnoslobodtsev},
  \citenamefont {Sabirianov}, \citenamefont {Sealy}, \citenamefont {Mei},
  \citenamefont {Wu},\ and\ \citenamefont {Tan}}]{Shen2021}%
  \BibitemOpen
  \bibfield  {author} {\bibinfo {author} {\bibfnamefont {X.}~\bibnamefont
  {Shen}}, \bibinfo {author} {\bibfnamefont {T.}~\bibnamefont {Sun}}, \bibinfo
  {author} {\bibfnamefont {L.}~\bibnamefont {Yang}}, \bibinfo {author}
  {\bibfnamefont {A.}~\bibnamefont {Krasnoslobodtsev}}, \bibinfo {author}
  {\bibfnamefont {R.}~\bibnamefont {Sabirianov}}, \bibinfo {author}
  {\bibfnamefont {M.}~\bibnamefont {Sealy}}, \bibinfo {author} {\bibfnamefont
  {W.-N.}\ \bibnamefont {Mei}}, \bibinfo {author} {\bibfnamefont
  {Z.}~\bibnamefont {Wu}},\ and\ \bibinfo {author} {\bibfnamefont
  {L.}~\bibnamefont {Tan}},\ }\bibfield  {title} {\enquote {\bibinfo {title}
  {Ultra-fast charging in aluminum-ion batteries: electric double layers on
  active anode},}\ }\href {https://doi.org/10.1038/s41467-021-21108-4}
  {\bibfield  {journal} {\bibinfo  {journal} {Nature Communications}\ }\textbf
  {\bibinfo {volume} {12}} (\bibinfo {year} {2021}),\
  10.1038/s41467-021-21108-4}\BibitemShut {NoStop}%
\bibitem [{\citenamefont {Safarpour}\ and\ \citenamefont
  {Khataee}(2019)}]{Safarpour}%
  \BibitemOpen
  \bibfield  {author} {\bibinfo {author} {\bibfnamefont {M.}~\bibnamefont
  {Safarpour}}\ and\ \bibinfo {author} {\bibfnamefont {A.}~\bibnamefont
  {Khataee}},\ }\bibfield  {title} {\enquote {\bibinfo {title} {Chapter 15 -
  graphene-based materials for water purification},}\ }in\ \href
  {https://doi.org/https://doi.org/10.1016/B978-0-12-813926-4.00021-5} {\emph
  {\bibinfo {booktitle} {Nanoscale Materials in Water Purification}}},\
  \bibinfo {series and number} {Micro and Nano Technologies},\ \bibinfo
  {editor} {edited by\ \bibinfo {editor} {\bibfnamefont {S.}~\bibnamefont
  {Thomas}}, \bibinfo {editor} {\bibfnamefont {D.}~\bibnamefont {Pasquini}},
  \bibinfo {editor} {\bibfnamefont {S.-Y.}\ \bibnamefont {Leu}},\ and\ \bibinfo
  {editor} {\bibfnamefont {D.~A.}\ \bibnamefont {Gopakumar}}}\ (\bibinfo
  {publisher} {Elsevier},\ \bibinfo {year} {2019})\ pp.\ \bibinfo {pages}
  {383--430}\BibitemShut {NoStop}%
\bibitem [{\citenamefont {Ray}, \citenamefont {Gusain},\ and\ \citenamefont
  {Kumar}(2020)}]{Ray}%
  \BibitemOpen
  \bibfield  {author} {\bibinfo {author} {\bibfnamefont {S.~S.}\ \bibnamefont
  {Ray}}, \bibinfo {author} {\bibfnamefont {R.}~\bibnamefont {Gusain}},\ and\
  \bibinfo {author} {\bibfnamefont {N.}~\bibnamefont {Kumar}},\ }\bibfield
  {title} {\enquote {\bibinfo {title} {Chapter ten - two-dimensional carbon
  nanomaterials-based adsorbents},}\ }in\ \href
  {https://doi.org/https://doi.org/10.1016/B978-0-12-821959-1.00010-6} {\emph
  {\bibinfo {booktitle} {Carbon Nanomaterial-Based Adsorbents for Water
  Purification}}},\ \bibinfo {series and number} {Micro and Nano
  Technologies},\ \bibinfo {editor} {edited by\ \bibinfo {editor}
  {\bibfnamefont {S.~S.}\ \bibnamefont {Ray}}, \bibinfo {editor} {\bibfnamefont
  {R.}~\bibnamefont {Gusain}},\ and\ \bibinfo {editor} {\bibfnamefont
  {N.}~\bibnamefont {Kumar}}}\ (\bibinfo  {publisher} {Elsevier},\ \bibinfo
  {year} {2020})\ pp.\ \bibinfo {pages} {225--273}\BibitemShut {NoStop}%
\bibitem [{\citenamefont {Han}\ \emph {et~al.}(1998)\citenamefont {Han},
  \citenamefont {Cho}, \citenamefont {Lee},\ and\ \citenamefont {Kim}}]{Han98}%
  \BibitemOpen
  \bibfield  {author} {\bibinfo {author} {\bibfnamefont {J.}~\bibnamefont
  {Han}}, \bibinfo {author} {\bibfnamefont {K.}~\bibnamefont {Cho}}, \bibinfo
  {author} {\bibfnamefont {K.-H.}\ \bibnamefont {Lee}},\ and\ \bibinfo {author}
  {\bibfnamefont {H.}~\bibnamefont {Kim}},\ }\bibfield  {title} {\enquote
  {\bibinfo {title} {Porous graphite matrix for chemical heat pumps},}\ }\href
  {https://doi.org/https://doi.org/10.1016/S0008-6223(98)00150-X} {\bibfield
  {journal} {\bibinfo  {journal} {Carbon}\ }\textbf {\bibinfo {volume} {36}},\
  \bibinfo {pages} {1801--1810} (\bibinfo {year} {1998})}\BibitemShut {NoStop}%
\bibitem [{\citenamefont {Inagaki}, \citenamefont {Qiu},\ and\ \citenamefont
  {Guo}(2015)}]{Inagaki15}%
  \BibitemOpen
  \bibfield  {author} {\bibinfo {author} {\bibfnamefont {M.}~\bibnamefont
  {Inagaki}}, \bibinfo {author} {\bibfnamefont {J.}~\bibnamefont {Qiu}},\ and\
  \bibinfo {author} {\bibfnamefont {Q.}~\bibnamefont {Guo}},\ }\bibfield
  {title} {\enquote {\bibinfo {title} {Carbon foam: Preparation and
  application},}\ }\href
  {https://doi.org/https://doi.org/10.1016/j.carbon.2015.02.021} {\bibfield
  {journal} {\bibinfo  {journal} {Carbon}\ }\textbf {\bibinfo {volume} {87}},\
  \bibinfo {pages} {128--152} (\bibinfo {year} {2015})}\BibitemShut {NoStop}%
\bibitem [{\citenamefont {Zhong}\ \emph {et~al.}(2010)\citenamefont {Zhong},
  \citenamefont {Li}, \citenamefont {Wei}, \citenamefont {Liu}, \citenamefont
  {Guo}, \citenamefont {Shi},\ and\ \citenamefont {Liu}}]{Zhong10}%
  \BibitemOpen
  \bibfield  {author} {\bibinfo {author} {\bibfnamefont {Y.}~\bibnamefont
  {Zhong}}, \bibinfo {author} {\bibfnamefont {S.}~\bibnamefont {Li}}, \bibinfo
  {author} {\bibfnamefont {X.}~\bibnamefont {Wei}}, \bibinfo {author}
  {\bibfnamefont {Z.}~\bibnamefont {Liu}}, \bibinfo {author} {\bibfnamefont
  {Q.}~\bibnamefont {Guo}}, \bibinfo {author} {\bibfnamefont {J.}~\bibnamefont
  {Shi}},\ and\ \bibinfo {author} {\bibfnamefont {L.}~\bibnamefont {Liu}},\
  }\bibfield  {title} {\enquote {\bibinfo {title} {Heat transfer enhancement of
  paraffin wax using compressed expanded natural graphite for thermal energy
  storage},}\ }\href
  {https://doi.org/https://doi.org/10.1016/j.carbon.2009.09.033} {\bibfield
  {journal} {\bibinfo  {journal} {Carbon}\ }\textbf {\bibinfo {volume} {48}},\
  \bibinfo {pages} {300--304} (\bibinfo {year} {2010})}\BibitemShut {NoStop}%
\bibitem [{\citenamefont {Paul}\ \emph {et~al.}(2012)\citenamefont {Paul},
  \citenamefont {Voevodin}, \citenamefont {Zemlyanov}, \citenamefont {Roy},\
  and\ \citenamefont {Fisher}}]{Rajib}%
  \BibitemOpen
  \bibfield  {author} {\bibinfo {author} {\bibfnamefont {R.}~\bibnamefont
  {Paul}}, \bibinfo {author} {\bibfnamefont {A.~A.}\ \bibnamefont {Voevodin}},
  \bibinfo {author} {\bibfnamefont {D.}~\bibnamefont {Zemlyanov}}, \bibinfo
  {author} {\bibfnamefont {A.~K.}\ \bibnamefont {Roy}},\ and\ \bibinfo {author}
  {\bibfnamefont {T.~S.}\ \bibnamefont {Fisher}},\ }\bibfield  {title}
  {\enquote {\bibinfo {title} {{Microwave-Assisted Surface Synthesis of a
  Boron–Carbon–Nitrogen Foam and its Desorption Enthalpy}},}\ }\href
  {https://doi.org/https://doi.org/10.1002/adfm.201200325} {\bibfield
  {journal} {\bibinfo  {journal} {Advanced Functional Materials}\ }\textbf
  {\bibinfo {volume} {22}},\ \bibinfo {pages} {3682--3690} (\bibinfo {year}
  {2012})},\ \Eprint
  {https://arxiv.org/abs/https://onlinelibrary.wiley.com/doi/pdf/10.1002/adfm.201200325}
  {https://onlinelibrary.wiley.com/doi/pdf/10.1002/adfm.201200325} \BibitemShut
  {NoStop}%
\bibitem [{\citenamefont {Gallego}\ and\ \citenamefont
  {Klett}(2003)}]{Gallego03}%
  \BibitemOpen
  \bibfield  {author} {\bibinfo {author} {\bibfnamefont {N.~C.}\ \bibnamefont
  {Gallego}}\ and\ \bibinfo {author} {\bibfnamefont {J.~W.}\ \bibnamefont
  {Klett}},\ }\bibfield  {title} {\enquote {\bibinfo {title} {Carbon foams for
  thermal management},}\ }\href
  {https://doi.org/https://doi.org/10.1016/S0008-6223(03)00091-5} {\bibfield
  {journal} {\bibinfo  {journal} {Carbon}\ }\textbf {\bibinfo {volume} {41}},\
  \bibinfo {pages} {1461--1466} (\bibinfo {year} {2003})}\BibitemShut {NoStop}%
\bibitem [{\citenamefont {Li}\ \emph {et~al.}(2012)\citenamefont {Li},
  \citenamefont {Chen}, \citenamefont {Ren}, \citenamefont {Li},\ and\
  \citenamefont {Cheng}}]{Li17}%
  \BibitemOpen
  \bibfield  {author} {\bibinfo {author} {\bibfnamefont {N.}~\bibnamefont
  {Li}}, \bibinfo {author} {\bibfnamefont {Z.}~\bibnamefont {Chen}}, \bibinfo
  {author} {\bibfnamefont {W.}~\bibnamefont {Ren}}, \bibinfo {author}
  {\bibfnamefont {F.}~\bibnamefont {Li}},\ and\ \bibinfo {author}
  {\bibfnamefont {H.-M.}\ \bibnamefont {Cheng}},\ }\bibfield  {title} {\enquote
  {\bibinfo {title} {Flexible graphene-based lithium ion batteries with
  ultrafast charge and discharge rates},}\ }\href
  {https://doi.org/10.1073/pnas.1210072109} {\bibfield  {journal} {\bibinfo
  {journal} {Proceedings of the National Academy of Sciences}\ }\textbf
  {\bibinfo {volume} {109}},\ \bibinfo {pages} {17360--17365} (\bibinfo {year}
  {2012})},\ \Eprint
  {https://arxiv.org/abs/https://www.pnas.org/content/109/43/17360.full.pdf}
  {https://www.pnas.org/content/109/43/17360.full.pdf} \BibitemShut {NoStop}%
\bibitem [{\citenamefont {Xue}\ \emph {et~al.}(2013)\citenamefont {Xue},
  \citenamefont {Yu}, \citenamefont {Dai}, \citenamefont {Wang}, \citenamefont
  {Li}, \citenamefont {Roy}, \citenamefont {Lu}, \citenamefont {Chen},
  \citenamefont {Liu},\ and\ \citenamefont {Qu}}]{Xue}%
  \BibitemOpen
  \bibfield  {author} {\bibinfo {author} {\bibfnamefont {Y.}~\bibnamefont
  {Xue}}, \bibinfo {author} {\bibfnamefont {D.}~\bibnamefont {Yu}}, \bibinfo
  {author} {\bibfnamefont {L.}~\bibnamefont {Dai}}, \bibinfo {author}
  {\bibfnamefont {R.}~\bibnamefont {Wang}}, \bibinfo {author} {\bibfnamefont
  {D.}~\bibnamefont {Li}}, \bibinfo {author} {\bibfnamefont {A.}~\bibnamefont
  {Roy}}, \bibinfo {author} {\bibfnamefont {F.}~\bibnamefont {Lu}}, \bibinfo
  {author} {\bibfnamefont {H.}~\bibnamefont {Chen}}, \bibinfo {author}
  {\bibfnamefont {Y.}~\bibnamefont {Liu}},\ and\ \bibinfo {author}
  {\bibfnamefont {J.}~\bibnamefont {Qu}},\ }\bibfield  {title} {\enquote
  {\bibinfo {title} {Three-dimensional b{,}n-doped graphene foam as a
  metal-free catalyst for oxygen reduction reaction},}\ }\href
  {https://doi.org/10.1039/C3CP51942B} {\bibfield  {journal} {\bibinfo
  {journal} {Phys. Chem. Chem. Phys.}\ }\textbf {\bibinfo {volume} {15}},\
  \bibinfo {pages} {12220--12226} (\bibinfo {year} {2013})}\BibitemShut
  {NoStop}%
\bibitem [{\citenamefont {Teich}\ \emph {et~al.}(2020)\citenamefont {Teich},
  \citenamefont {Dvir}, \citenamefont {Henning}, \citenamefont {Hamo},
  \citenamefont {Moody}, \citenamefont {Jurca}, \citenamefont {Cohen},
  \citenamefont {Marks}, \citenamefont {Rosen}, \citenamefont {Lauhon},\ and\
  \citenamefont {Ismach}}]{Teich}%
  \BibitemOpen
  \bibfield  {author} {\bibinfo {author} {\bibfnamefont {J.}~\bibnamefont
  {Teich}}, \bibinfo {author} {\bibfnamefont {R.}~\bibnamefont {Dvir}},
  \bibinfo {author} {\bibfnamefont {A.}~\bibnamefont {Henning}}, \bibinfo
  {author} {\bibfnamefont {E.~R.}\ \bibnamefont {Hamo}}, \bibinfo {author}
  {\bibfnamefont {M.~J.}\ \bibnamefont {Moody}}, \bibinfo {author}
  {\bibfnamefont {T.}~\bibnamefont {Jurca}}, \bibinfo {author} {\bibfnamefont
  {H.}~\bibnamefont {Cohen}}, \bibinfo {author} {\bibfnamefont {T.~J.}\
  \bibnamefont {Marks}}, \bibinfo {author} {\bibfnamefont {B.~A.}\ \bibnamefont
  {Rosen}}, \bibinfo {author} {\bibfnamefont {L.~J.}\ \bibnamefont {Lauhon}},\
  and\ \bibinfo {author} {\bibfnamefont {A.}~\bibnamefont {Ismach}},\
  }\bibfield  {title} {\enquote {\bibinfo {title} {{Light and complex 3D
  MoS$_2$/graphene heterostructures as efficient catalysts for the hydrogen
  evolution reaction}},}\ }\href {https://doi.org/10.1039/C9NR09564K}
  {\bibfield  {journal} {\bibinfo  {journal} {Nanoscale}\ }\textbf {\bibinfo
  {volume} {12}},\ \bibinfo {pages} {2715--2725} (\bibinfo {year}
  {2020})}\BibitemShut {NoStop}%
\bibitem [{\citenamefont {Zhang}\ \emph {et~al.}(2016)\citenamefont {Zhang},
  \citenamefont {DeArmond}, \citenamefont {Alvarez}, \citenamefont {Zhao},
  \citenamefont {Wang}, \citenamefont {Hou}, \citenamefont {Malik},
  \citenamefont {Heineman},\ and\ \citenamefont {Shanov}}]{Zhang16}%
  \BibitemOpen
  \bibfield  {author} {\bibinfo {author} {\bibfnamefont {L.}~\bibnamefont
  {Zhang}}, \bibinfo {author} {\bibfnamefont {D.}~\bibnamefont {DeArmond}},
  \bibinfo {author} {\bibfnamefont {N.~T.}\ \bibnamefont {Alvarez}}, \bibinfo
  {author} {\bibfnamefont {D.}~\bibnamefont {Zhao}}, \bibinfo {author}
  {\bibfnamefont {T.}~\bibnamefont {Wang}}, \bibinfo {author} {\bibfnamefont
  {G.}~\bibnamefont {Hou}}, \bibinfo {author} {\bibfnamefont {R.}~\bibnamefont
  {Malik}}, \bibinfo {author} {\bibfnamefont {W.~R.}\ \bibnamefont
  {Heineman}},\ and\ \bibinfo {author} {\bibfnamefont {V.}~\bibnamefont
  {Shanov}},\ }\bibfield  {title} {\enquote {\bibinfo {title} {Beyond graphene
  foam{,} a new form of three-dimensional graphene for supercapacitor
  electrodes},}\ }\href {https://doi.org/10.1039/C5TA10031C} {\bibfield
  {journal} {\bibinfo  {journal} {J. Mater. Chem. A}\ }\textbf {\bibinfo
  {volume} {4}},\ \bibinfo {pages} {1876--1886} (\bibinfo {year}
  {2016})}\BibitemShut {NoStop}%
\bibitem [{\citenamefont {Sun}\ \emph {et~al.}(2020)\citenamefont {Sun},
  \citenamefont {Liu}, \citenamefont {Sheng}, \citenamefont {Huang},
  \citenamefont {Lv}, \citenamefont {Zhou},\ and\ \citenamefont
  {Cheng}}]{D0MH00815J}%
  \BibitemOpen
  \bibfield  {author} {\bibinfo {author} {\bibfnamefont {C.}~\bibnamefont
  {Sun}}, \bibinfo {author} {\bibfnamefont {Y.}~\bibnamefont {Liu}}, \bibinfo
  {author} {\bibfnamefont {J.}~\bibnamefont {Sheng}}, \bibinfo {author}
  {\bibfnamefont {Q.}~\bibnamefont {Huang}}, \bibinfo {author} {\bibfnamefont
  {W.}~\bibnamefont {Lv}}, \bibinfo {author} {\bibfnamefont {G.}~\bibnamefont
  {Zhou}},\ and\ \bibinfo {author} {\bibfnamefont {H.-M.}\ \bibnamefont
  {Cheng}},\ }\bibfield  {title} {\enquote {\bibinfo {title} {Status and
  prospects of porous graphene networks for lithium–sulfur batteries},}\
  }\href {https://doi.org/10.1039/D0MH00815J} {\bibfield  {journal} {\bibinfo
  {journal} {Mater. Horiz.}\ }\textbf {\bibinfo {volume} {7}},\ \bibinfo
  {pages} {2487--2518} (\bibinfo {year} {2020})}\BibitemShut {NoStop}%
\bibitem [{\citenamefont {Dimitrakakis}, \citenamefont {Tylianakis},\ and\
  \citenamefont {Froudakis}(2008)}]{Dimitrakakis}%
  \BibitemOpen
  \bibfield  {author} {\bibinfo {author} {\bibfnamefont {G.~K.}\ \bibnamefont
  {Dimitrakakis}}, \bibinfo {author} {\bibfnamefont {E.}~\bibnamefont
  {Tylianakis}},\ and\ \bibinfo {author} {\bibfnamefont {G.~E.}\ \bibnamefont
  {Froudakis}},\ }\bibfield  {title} {\enquote {\bibinfo {title} {{Pillared
  Graphene: A New 3-D Network Nanostructure for Enhanced Hydrogen Storage}},}\
  }\href {https://doi.org/10.1021/nl801417w} {\bibfield  {journal} {\bibinfo
  {journal} {Nano Letters}\ }\textbf {\bibinfo {volume} {8}},\ \bibinfo {pages}
  {3166--3170} (\bibinfo {year} {2008})},\ \Eprint
  {https://arxiv.org/abs/https://doi.org/10.1021/nl801417w}
  {https://doi.org/10.1021/nl801417w} \BibitemShut {NoStop}%
\bibitem [{\citenamefont {Duan}\ \emph {et~al.}(2017)\citenamefont {Duan},
  \citenamefont {Li}, \citenamefont {Hu},\ and\ \citenamefont {Wang}}]{Duan}%
  \BibitemOpen
  \bibfield  {author} {\bibinfo {author} {\bibfnamefont {K.}~\bibnamefont
  {Duan}}, \bibinfo {author} {\bibfnamefont {L.}~\bibnamefont {Li}}, \bibinfo
  {author} {\bibfnamefont {Y.}~\bibnamefont {Hu}},\ and\ \bibinfo {author}
  {\bibfnamefont {X.}~\bibnamefont {Wang}},\ }\bibfield  {title} {\enquote
  {\bibinfo {title} {Pillared graphene as an ultra-high sensitivity mass
  sensor},}\ }\href {https://doi.org/10.1038/s41598-017-14182-6} {\bibfield
  {journal} {\bibinfo  {journal} {Scientific Reports}\ }\textbf {\bibinfo
  {volume} {7}},\ \bibinfo {pages} {14012} (\bibinfo {year} {2017})},\ \Eprint
  {https://arxiv.org/abs/https://doi.org/10.1038/s41598-017-14182-6}
  {https://doi.org/10.1038/s41598-017-14182-6} \BibitemShut {NoStop}%
\bibitem [{\citenamefont {Leitgeb}\ \emph
  {et~al.}(2017{\natexlab{a}})\citenamefont {Leitgeb}, \citenamefont {Zellner},
  \citenamefont {Schneider}, \citenamefont {Schwab}, \citenamefont {Hutter},\
  and\ \citenamefont {Schmid}}]{Leitgeb}%
  \BibitemOpen
  \bibfield  {author} {\bibinfo {author} {\bibfnamefont {M.}~\bibnamefont
  {Leitgeb}}, \bibinfo {author} {\bibfnamefont {C.}~\bibnamefont {Zellner}},
  \bibinfo {author} {\bibfnamefont {M.}~\bibnamefont {Schneider}}, \bibinfo
  {author} {\bibfnamefont {S.}~\bibnamefont {Schwab}}, \bibinfo {author}
  {\bibfnamefont {H.}~\bibnamefont {Hutter}},\ and\ \bibinfo {author}
  {\bibfnamefont {U.}~\bibnamefont {Schmid}},\ }\bibfield  {title} {\enquote
  {\bibinfo {title} {{Metal assisted photochemical etching of 4H silicon
  carbide}},}\ }\href
  {https://doi.org/https://doi.org/10.1088/1361-6463/aa8942} {\bibfield
  {journal} {\bibinfo  {journal} {J. Phys. D: Appl. Phys.}\ }\textbf {\bibinfo
  {volume} {50}},\ \bibinfo {pages} {435301} (\bibinfo {year}
  {2017}{\natexlab{a}})}\BibitemShut {NoStop}%
\bibitem [{\citenamefont {Shor}\ \emph {et~al.}(1993)\citenamefont {Shor},
  \citenamefont {Grimberg}, \citenamefont {Weiss},\ and\ \citenamefont
  {Kurtz}}]{Shor1993}%
  \BibitemOpen
  \bibfield  {author} {\bibinfo {author} {\bibfnamefont {J.~S.}\ \bibnamefont
  {Shor}}, \bibinfo {author} {\bibfnamefont {I.}~\bibnamefont {Grimberg}},
  \bibinfo {author} {\bibfnamefont {B.-Z.}\ \bibnamefont {Weiss}},\ and\
  \bibinfo {author} {\bibfnamefont {A.~D.}\ \bibnamefont {Kurtz}},\ }\bibfield
  {title} {\enquote {\bibinfo {title} {Direct observation of porous {SiC}
  formed by anodization in {HF}},}\ }\href {https://doi.org/10.1063/1.109226}
  {\bibfield  {journal} {\bibinfo  {journal} {Applied Physics Letters}\
  }\textbf {\bibinfo {volume} {62}},\ \bibinfo {pages} {2836--2838} (\bibinfo
  {year} {1993})}\BibitemShut {NoStop}%
\bibitem [{\citenamefont {Leitgeb}\ \emph
  {et~al.}(2017{\natexlab{b}})\citenamefont {Leitgeb}, \citenamefont {Zellner},
  \citenamefont {Pfusterschmied}, \citenamefont {Schneider},\ and\
  \citenamefont {Schmid}}]{Leitgeb2017a}%
  \BibitemOpen
  \bibfield  {author} {\bibinfo {author} {\bibfnamefont {M.}~\bibnamefont
  {Leitgeb}}, \bibinfo {author} {\bibfnamefont {C.}~\bibnamefont {Zellner}},
  \bibinfo {author} {\bibfnamefont {G.}~\bibnamefont {Pfusterschmied}},
  \bibinfo {author} {\bibfnamefont {M.}~\bibnamefont {Schneider}},\ and\
  \bibinfo {author} {\bibfnamefont {U.}~\bibnamefont {Schmid}},\ }\bibfield
  {title} {\enquote {\bibinfo {title} {{Porous Silicon Carbide for {MEMS}}},}\
  }\href {https://doi.org/10.3390/proceedings1040297} {\bibfield  {journal}
  {\bibinfo  {journal} {Proceedings}\ }\textbf {\bibinfo {volume} {1}},\
  \bibinfo {pages} {297} (\bibinfo {year} {2017}{\natexlab{b}})}\BibitemShut
  {NoStop}%
\bibitem [{\citenamefont {Leitgeb}\ \emph
  {et~al.}(2017{\natexlab{c}})\citenamefont {Leitgeb}, \citenamefont {Zellner},
  \citenamefont {Schneider},\ and\ \citenamefont {Schmid}}]{Leitgeb2017}%
  \BibitemOpen
  \bibfield  {author} {\bibinfo {author} {\bibfnamefont {M.}~\bibnamefont
  {Leitgeb}}, \bibinfo {author} {\bibfnamefont {C.}~\bibnamefont {Zellner}},
  \bibinfo {author} {\bibfnamefont {M.}~\bibnamefont {Schneider}},\ and\
  \bibinfo {author} {\bibfnamefont {U.}~\bibnamefont {Schmid}},\ }\bibfield
  {title} {\enquote {\bibinfo {title} {Porous single crystalline 4h silicon
  carbide rugate mirrors},}\ }\href {https://doi.org/10.1063/1.5001876}
  {\bibfield  {journal} {\bibinfo  {journal} {{APL} Materials}\ }\textbf
  {\bibinfo {volume} {5}},\ \bibinfo {pages} {106106} (\bibinfo {year}
  {2017}{\natexlab{c}})}\BibitemShut {NoStop}%
\bibitem [{\citenamefont {Leitgeb}\ \emph {et~al.}(2018)\citenamefont
  {Leitgeb}, \citenamefont {Zellner}, \citenamefont {Schneider}, \citenamefont
  {Lukschanderl},\ and\ \citenamefont {Schmid}}]{Leitgeb1}%
  \BibitemOpen
  \bibfield  {author} {\bibinfo {author} {\bibfnamefont {M.}~\bibnamefont
  {Leitgeb}}, \bibinfo {author} {\bibfnamefont {C.}~\bibnamefont {Zellner}},
  \bibinfo {author} {\bibfnamefont {M.}~\bibnamefont {Schneider}}, \bibinfo
  {author} {\bibfnamefont {M.}~\bibnamefont {Lukschanderl}},\ and\ \bibinfo
  {author} {\bibfnamefont {U.}~\bibnamefont {Schmid}},\ }\bibfield  {title}
  {\enquote {\bibinfo {title} {A cellular automaton based interpretation of
  metal assisted photochemical porosification of 4h silicon carbide},}\ }\href
  {https://doi.org/https://doi.org/10.1149/2.0571809jes} {\bibfield  {journal}
  {\bibinfo  {journal} {J. Electrochem. Soc.}\ }\textbf {\bibinfo {volume}
  {165}},\ \bibinfo {pages} {E325} (\bibinfo {year} {2018})}\BibitemShut
  {NoStop}%
\bibitem [{\citenamefont {Leitgeb}\ \emph
  {et~al.}(2017{\natexlab{d}})\citenamefont {Leitgeb}, \citenamefont {Zellner},
  \citenamefont {Hufnagl}, \citenamefont {Schneider}, \citenamefont {Schwab},
  \citenamefont {Hutter},\ and\ \citenamefont {Schmid}}]{Leitgeb3}%
  \BibitemOpen
  \bibfield  {author} {\bibinfo {author} {\bibfnamefont {M.}~\bibnamefont
  {Leitgeb}}, \bibinfo {author} {\bibfnamefont {C.}~\bibnamefont {Zellner}},
  \bibinfo {author} {\bibfnamefont {C.}~\bibnamefont {Hufnagl}}, \bibinfo
  {author} {\bibfnamefont {M.}~\bibnamefont {Schneider}}, \bibinfo {author}
  {\bibfnamefont {S.}~\bibnamefont {Schwab}}, \bibinfo {author} {\bibfnamefont
  {H.}~\bibnamefont {Hutter}},\ and\ \bibinfo {author} {\bibfnamefont
  {U.}~\bibnamefont {Schmid}},\ }\bibfield  {title} {\enquote {\bibinfo {title}
  {Stacked layers of different porosity in 4h sic substrates applying a
  photoelectrochemical approach},}\ }\href
  {https://doi.org/https://doi.org/10.1149/2.0571809jes} {\bibfield  {journal}
  {\bibinfo  {journal} {J. Electrochem. Soc.}\ }\textbf {\bibinfo {volume}
  {164}},\ \bibinfo {pages} {E337} (\bibinfo {year}
  {2017}{\natexlab{d}})}\BibitemShut {NoStop}%
\bibitem [{\citenamefont {Leitgeb}\ \emph {et~al.}(2016)\citenamefont
  {Leitgeb}, \citenamefont {Zellner}, \citenamefont {Schneider},\ and\
  \citenamefont {Schmid}}]{Leitgeb2}%
  \BibitemOpen
  \bibfield  {author} {\bibinfo {author} {\bibfnamefont {M.}~\bibnamefont
  {Leitgeb}}, \bibinfo {author} {\bibfnamefont {C.}~\bibnamefont {Zellner}},
  \bibinfo {author} {\bibfnamefont {M.}~\bibnamefont {Schneider}},\ and\
  \bibinfo {author} {\bibfnamefont {U.}~\bibnamefont {Schmid}},\ }\bibfield
  {title} {\enquote {\bibinfo {title} {A combination of metal assisted
  photochemical and photoelectrochemical etching for tailored porosification of
  4h sic substrates},}\ }\href
  {https://doi.org/https://doi.org/10.1149/2.0041610jss} {\bibfield  {journal}
  {\bibinfo  {journal} {ECS J. Solid State Sci. Technol.}\ }\textbf {\bibinfo
  {volume} {5}},\ \bibinfo {pages} {P556} (\bibinfo {year} {2016})}\BibitemShut
  {NoStop}%
\bibitem [{\citenamefont {van Aarle}\ \emph {et~al.}(2015)\citenamefont {van
  Aarle}, \citenamefont {Palenstijn}, \citenamefont {Beenhouwer}, \citenamefont
  {Altantzis}, \citenamefont {Bals}, \citenamefont {Batenburg},\ and\
  \citenamefont {Sijbers}}]{vanAarle2015}%
  \BibitemOpen
  \bibfield  {author} {\bibinfo {author} {\bibfnamefont {W.}~\bibnamefont {van
  Aarle}}, \bibinfo {author} {\bibfnamefont {W.~J.}\ \bibnamefont
  {Palenstijn}}, \bibinfo {author} {\bibfnamefont {J.~D.}\ \bibnamefont
  {Beenhouwer}}, \bibinfo {author} {\bibfnamefont {T.}~\bibnamefont
  {Altantzis}}, \bibinfo {author} {\bibfnamefont {S.}~\bibnamefont {Bals}},
  \bibinfo {author} {\bibfnamefont {K.~J.}\ \bibnamefont {Batenburg}},\ and\
  \bibinfo {author} {\bibfnamefont {J.}~\bibnamefont {Sijbers}},\ }\bibfield
  {title} {\enquote {\bibinfo {title} {The {ASTRA} toolbox: A platform for
  advanced algorithm development in electron tomography},}\ }\href
  {https://doi.org/10.1016/j.ultramic.2015.05.002} {\bibfield  {journal}
  {\bibinfo  {journal} {Ultramicroscopy}\ }\textbf {\bibinfo {volume} {157}},\
  \bibinfo {pages} {35--47} (\bibinfo {year} {2015})}\BibitemShut {NoStop}%
\bibitem [{\citenamefont {Ferrari}\ \emph {et~al.}(2006)\citenamefont
  {Ferrari}, \citenamefont {Meyer}, \citenamefont {Scardaci}, \citenamefont
  {Casiraghi}, \citenamefont {Lazzeri}, \citenamefont {Mauri}, \citenamefont
  {Piscanec}, \citenamefont {Jiang}, \citenamefont {Novoselov}, \citenamefont
  {Roth} \emph {et~al.}}]{Ferrari06}%
  \BibitemOpen
  \bibfield  {author} {\bibinfo {author} {\bibfnamefont {A.~C.}\ \bibnamefont
  {Ferrari}}, \bibinfo {author} {\bibfnamefont {J.}~\bibnamefont {Meyer}},
  \bibinfo {author} {\bibfnamefont {V.}~\bibnamefont {Scardaci}}, \bibinfo
  {author} {\bibfnamefont {C.}~\bibnamefont {Casiraghi}}, \bibinfo {author}
  {\bibfnamefont {M.}~\bibnamefont {Lazzeri}}, \bibinfo {author} {\bibfnamefont
  {F.}~\bibnamefont {Mauri}}, \bibinfo {author} {\bibfnamefont
  {S.}~\bibnamefont {Piscanec}}, \bibinfo {author} {\bibfnamefont
  {D.}~\bibnamefont {Jiang}}, \bibinfo {author} {\bibfnamefont
  {K.}~\bibnamefont {Novoselov}}, \bibinfo {author} {\bibfnamefont
  {S.}~\bibnamefont {Roth}}, \emph {et~al.},\ }\bibfield  {title} {\enquote
  {\bibinfo {title} {Raman spectrum of graphene and graphene layers},}\
  }\href@noop {} {\bibfield  {journal} {\bibinfo  {journal} {Physical Review
  Letters}\ }\textbf {\bibinfo {volume} {97}},\ \bibinfo {pages} {187401}
  (\bibinfo {year} {2006})}\BibitemShut {NoStop}%
\bibitem [{\citenamefont {Malard}\ \emph {et~al.}(2009)\citenamefont {Malard},
  \citenamefont {Pimenta}, \citenamefont {Dresselhaus},\ and\ \citenamefont
  {Dresselhaus}}]{Malard09}%
  \BibitemOpen
  \bibfield  {author} {\bibinfo {author} {\bibfnamefont {L.~M.}\ \bibnamefont
  {Malard}}, \bibinfo {author} {\bibfnamefont {M.~A.}\ \bibnamefont {Pimenta}},
  \bibinfo {author} {\bibfnamefont {G.}~\bibnamefont {Dresselhaus}},\ and\
  \bibinfo {author} {\bibfnamefont {M.~S.}\ \bibnamefont {Dresselhaus}},\
  }\bibfield  {title} {\enquote {\bibinfo {title} {Raman spectroscopy in
  graphene},}\ }\href
  {https://doi.org/https://doi.org/10.1016/j.physrep.2009.02.003} {\bibfield
  {journal} {\bibinfo  {journal} {Physics Reports}\ }\textbf {\bibinfo {volume}
  {473}},\ \bibinfo {pages} {51--87} (\bibinfo {year} {2009})}\BibitemShut
  {NoStop}%
\bibitem [{\citenamefont {Eckmann}\ \emph {et~al.}(2012)\citenamefont
  {Eckmann}, \citenamefont {Felten}, \citenamefont {Mishchenko}, \citenamefont
  {Britnell}, \citenamefont {Krupke}, \citenamefont {Novoselov},\ and\
  \citenamefont {Casiraghi}}]{Eckmann12}%
  \BibitemOpen
  \bibfield  {author} {\bibinfo {author} {\bibfnamefont {A.}~\bibnamefont
  {Eckmann}}, \bibinfo {author} {\bibfnamefont {A.}~\bibnamefont {Felten}},
  \bibinfo {author} {\bibfnamefont {A.}~\bibnamefont {Mishchenko}}, \bibinfo
  {author} {\bibfnamefont {L.}~\bibnamefont {Britnell}}, \bibinfo {author}
  {\bibfnamefont {R.}~\bibnamefont {Krupke}}, \bibinfo {author} {\bibfnamefont
  {K.~S.}\ \bibnamefont {Novoselov}},\ and\ \bibinfo {author} {\bibfnamefont
  {C.}~\bibnamefont {Casiraghi}},\ }\bibfield  {title} {\enquote {\bibinfo
  {title} {Probing the nature of defects in graphene by raman spectroscopy},}\
  }\href@noop {} {\bibfield  {journal} {\bibinfo  {journal} {Nano Letters}\
  }\textbf {\bibinfo {volume} {12}},\ \bibinfo {pages} {3925--3930} (\bibinfo
  {year} {2012})}\BibitemShut {NoStop}%
\bibitem [{\citenamefont {Hähnlein}\ \emph {et~al.}(2020)\citenamefont
  {Hähnlein}, \citenamefont {Lebedev}, \citenamefont {Eliseyev}, \citenamefont
  {Smirnov}, \citenamefont {Davydov}, \citenamefont {Zubov}, \citenamefont
  {Lebedev},\ and\ \citenamefont {Pezoldt}}]{Hahnlein}%
  \BibitemOpen
  \bibfield  {author} {\bibinfo {author} {\bibfnamefont {B.}~\bibnamefont
  {Hähnlein}}, \bibinfo {author} {\bibfnamefont {S.}~\bibnamefont {Lebedev}},
  \bibinfo {author} {\bibfnamefont {I.}~\bibnamefont {Eliseyev}}, \bibinfo
  {author} {\bibfnamefont {A.}~\bibnamefont {Smirnov}}, \bibinfo {author}
  {\bibfnamefont {V.}~\bibnamefont {Davydov}}, \bibinfo {author} {\bibfnamefont
  {A.}~\bibnamefont {Zubov}}, \bibinfo {author} {\bibfnamefont
  {A.}~\bibnamefont {Lebedev}},\ and\ \bibinfo {author} {\bibfnamefont
  {J.}~\bibnamefont {Pezoldt}},\ }\bibfield  {title} {\enquote {\bibinfo
  {title} {Investigation of epitaxial graphene via raman spectroscopy: Origins
  of phonon mode asymmetries and line width deviations},}\ }\href
  {https://doi.org/https://doi.org/10.1016/j.carbon.2020.07.016} {\bibfield
  {journal} {\bibinfo  {journal} {Carbon}\ }\textbf {\bibinfo {volume} {170}},\
  \bibinfo {pages} {666--676} (\bibinfo {year} {2020})}\BibitemShut {NoStop}%
\bibitem [{\citenamefont {Das}\ \emph {et~al.}(2009)\citenamefont {Das},
  \citenamefont {Chakraborty}, \citenamefont {Piscanec}, \citenamefont
  {Pisana}, \citenamefont {Sood},\ and\ \citenamefont {Ferrari}}]{Ferrari09}%
  \BibitemOpen
  \bibfield  {author} {\bibinfo {author} {\bibfnamefont {A.}~\bibnamefont
  {Das}}, \bibinfo {author} {\bibfnamefont {B.}~\bibnamefont {Chakraborty}},
  \bibinfo {author} {\bibfnamefont {S.}~\bibnamefont {Piscanec}}, \bibinfo
  {author} {\bibfnamefont {S.}~\bibnamefont {Pisana}}, \bibinfo {author}
  {\bibfnamefont {A.~K.}\ \bibnamefont {Sood}},\ and\ \bibinfo {author}
  {\bibfnamefont {A.~C.}\ \bibnamefont {Ferrari}},\ }\bibfield  {title}
  {\enquote {\bibinfo {title} {Phonon renormalization in doped bilayer
  graphene},}\ }\href {https://doi.org/10.1103/PhysRevB.79.155417} {\bibfield
  {journal} {\bibinfo  {journal} {Phys. Rev. B}\ }\textbf {\bibinfo {volume}
  {79}},\ \bibinfo {pages} {155417} (\bibinfo {year} {2009})}\BibitemShut
  {NoStop}%
\bibitem [{\citenamefont {Das}\ \emph {et~al.}(2008)\citenamefont {Das},
  \citenamefont {Pisana}, \citenamefont {Chakraborty}, \citenamefont
  {Piscanec}, \citenamefont {Saha}, \citenamefont {Waghmare}, \citenamefont
  {Novoselov}, \citenamefont {Krishnamurthy}, \citenamefont {Geim},
  \citenamefont {Ferrari},\ and\ \citenamefont {Sood}}]{Das}%
  \BibitemOpen
  \bibfield  {author} {\bibinfo {author} {\bibfnamefont {A.}~\bibnamefont
  {Das}}, \bibinfo {author} {\bibfnamefont {S.}~\bibnamefont {Pisana}},
  \bibinfo {author} {\bibfnamefont {B.}~\bibnamefont {Chakraborty}}, \bibinfo
  {author} {\bibfnamefont {S.}~\bibnamefont {Piscanec}}, \bibinfo {author}
  {\bibfnamefont {S.~K.}\ \bibnamefont {Saha}}, \bibinfo {author}
  {\bibfnamefont {U.~V.}\ \bibnamefont {Waghmare}}, \bibinfo {author}
  {\bibfnamefont {K.~S.}\ \bibnamefont {Novoselov}}, \bibinfo {author}
  {\bibfnamefont {H.~R.}\ \bibnamefont {Krishnamurthy}}, \bibinfo {author}
  {\bibfnamefont {A.~K.}\ \bibnamefont {Geim}}, \bibinfo {author}
  {\bibfnamefont {A.~C.}\ \bibnamefont {Ferrari}},\ and\ \bibinfo {author}
  {\bibfnamefont {A.~K.}\ \bibnamefont {Sood}},\ }\bibfield  {title} {\enquote
  {\bibinfo {title} {{Monitoring dopants by Raman scattering in an
  electrochemically top-gated graphene transistor}},}\ }\href
  {https://doi.org/10.1038/nnano.2008.67} {\bibfield  {journal} {\bibinfo
  {journal} {Nature Nanotech.}\ }\textbf {\bibinfo {volume} {3}},\ \bibinfo
  {pages} {210--215} (\bibinfo {year} {2008})}\BibitemShut {NoStop}%
\bibitem [{\citenamefont {Lee}\ \emph {et~al.}(2008)\citenamefont {Lee},
  \citenamefont {Riedl}, \citenamefont {Krauss}, \citenamefont {von Klitzing},
  \citenamefont {Starke},\ and\ \citenamefont {Smet}}]{Lee}%
  \BibitemOpen
  \bibfield  {author} {\bibinfo {author} {\bibfnamefont {D.~S.}\ \bibnamefont
  {Lee}}, \bibinfo {author} {\bibfnamefont {C.}~\bibnamefont {Riedl}}, \bibinfo
  {author} {\bibfnamefont {B.}~\bibnamefont {Krauss}}, \bibinfo {author}
  {\bibfnamefont {K.}~\bibnamefont {von Klitzing}}, \bibinfo {author}
  {\bibfnamefont {U.}~\bibnamefont {Starke}},\ and\ \bibinfo {author}
  {\bibfnamefont {J.~H.}\ \bibnamefont {Smet}},\ }\bibfield  {title} {\enquote
  {\bibinfo {title} {{Raman Spectra of Epitaxial Graphene on SiC and of
  Epitaxial Graphene Transferred to SiO$_2$}},}\ }\href
  {https://doi.org/10.1021/nl802156w} {\bibfield  {journal} {\bibinfo
  {journal} {Nano Letters}\ }\textbf {\bibinfo {volume} {8}},\ \bibinfo {pages}
  {4320--4325} (\bibinfo {year} {2008})},\ \Eprint
  {https://arxiv.org/abs/https://doi.org/10.1021/nl802156w}
  {https://doi.org/10.1021/nl802156w} \BibitemShut {NoStop}%
\bibitem [{\citenamefont {Neumann}\ \emph {et~al.}(2015)\citenamefont
  {Neumann}, \citenamefont {Reichardt}, \citenamefont {Venezuela},
  \citenamefont {Dr\"{o}geler}, \citenamefont {Banszerus}, \citenamefont
  {Schmitz}, \citenamefont {Watanabe}, \citenamefont {Taniguchi}, \citenamefont
  {Mauri}, \citenamefont {Beschoten}, \citenamefont {Rotkin},\ and\
  \citenamefont {Stampfer}}]{Neumann}%
  \BibitemOpen
  \bibfield  {author} {\bibinfo {author} {\bibfnamefont {C.}~\bibnamefont
  {Neumann}}, \bibinfo {author} {\bibfnamefont {S.}~\bibnamefont {Reichardt}},
  \bibinfo {author} {\bibfnamefont {P.}~\bibnamefont {Venezuela}}, \bibinfo
  {author} {\bibfnamefont {M.}~\bibnamefont {Dr\"{o}geler}}, \bibinfo {author}
  {\bibfnamefont {L.}~\bibnamefont {Banszerus}}, \bibinfo {author}
  {\bibfnamefont {M.}~\bibnamefont {Schmitz}}, \bibinfo {author} {\bibfnamefont
  {K.}~\bibnamefont {Watanabe}}, \bibinfo {author} {\bibfnamefont
  {T.}~\bibnamefont {Taniguchi}}, \bibinfo {author} {\bibfnamefont
  {F.}~\bibnamefont {Mauri}}, \bibinfo {author} {\bibfnamefont
  {B.}~\bibnamefont {Beschoten}}, \bibinfo {author} {\bibfnamefont {S.~V.}\
  \bibnamefont {Rotkin}},\ and\ \bibinfo {author} {\bibfnamefont
  {C.}~\bibnamefont {Stampfer}},\ }\bibfield  {title} {\enquote {\bibinfo
  {title} {Raman spectroscopy as probe of nanometre-scale strain variations in
  graphene},}\ }\href {https://doi.org/10.1038/ncomms9429} {\bibfield
  {journal} {\bibinfo  {journal} {Nature Communications}\ }\textbf {\bibinfo
  {volume} {6}},\ \bibinfo {pages} {8429} (\bibinfo {year} {2015})}\BibitemShut
  {NoStop}%
\bibitem [{\citenamefont {Can{\c{c}}ado}\ \emph {et~al.}(2011)\citenamefont
  {Can{\c{c}}ado}, \citenamefont {Jorio}, \citenamefont {Ferreira},
  \citenamefont {Stavale}, \citenamefont {Achete}, \citenamefont {Capaz},
  \citenamefont {Moutinho}, \citenamefont {Lombardo}, \citenamefont {Kulmala},\
  and\ \citenamefont {Ferrari}}]{Canado11}%
  \BibitemOpen
  \bibfield  {author} {\bibinfo {author} {\bibfnamefont {L.~G.}\ \bibnamefont
  {Can{\c{c}}ado}}, \bibinfo {author} {\bibfnamefont {A.}~\bibnamefont
  {Jorio}}, \bibinfo {author} {\bibfnamefont {E.~H.~M.}\ \bibnamefont
  {Ferreira}}, \bibinfo {author} {\bibfnamefont {F.}~\bibnamefont {Stavale}},
  \bibinfo {author} {\bibfnamefont {C.~A.}\ \bibnamefont {Achete}}, \bibinfo
  {author} {\bibfnamefont {R.~B.}\ \bibnamefont {Capaz}}, \bibinfo {author}
  {\bibfnamefont {M.~V.~O.}\ \bibnamefont {Moutinho}}, \bibinfo {author}
  {\bibfnamefont {A.}~\bibnamefont {Lombardo}}, \bibinfo {author}
  {\bibfnamefont {T.~S.}\ \bibnamefont {Kulmala}},\ and\ \bibinfo {author}
  {\bibfnamefont {A.~C.}\ \bibnamefont {Ferrari}},\ }\bibfield  {title}
  {\enquote {\bibinfo {title} {Quantifying defects in graphene via raman
  spectroscopy at different excitation energies},}\ }\href
  {https://doi.org/10.1021/nl201432g} {\bibfield  {journal} {\bibinfo
  {journal} {Nano Letters}\ }\textbf {\bibinfo {volume} {11}},\ \bibinfo
  {pages} {3190--3196} (\bibinfo {year} {2011})}\BibitemShut {NoStop}%
\bibitem [{\citenamefont {Ribeiro-Soares}\ \emph {et~al.}(2015)\citenamefont
  {Ribeiro-Soares}, \citenamefont {Oliveros}, \citenamefont {Garin},
  \citenamefont {David}, \citenamefont {Martins}, \citenamefont {Almeida},
  \citenamefont {Martins-Ferreira}, \citenamefont {Takai}, \citenamefont
  {Enoki}, \citenamefont {Magalhães-Paniago}, \citenamefont {Malachias},
  \citenamefont {Jorio}, \citenamefont {Archanjo}, \citenamefont {Achete},\
  and\ \citenamefont {Cançado}}]{Ribeiro}%
  \BibitemOpen
  \bibfield  {author} {\bibinfo {author} {\bibfnamefont {J.}~\bibnamefont
  {Ribeiro-Soares}}, \bibinfo {author} {\bibfnamefont {M.}~\bibnamefont
  {Oliveros}}, \bibinfo {author} {\bibfnamefont {C.}~\bibnamefont {Garin}},
  \bibinfo {author} {\bibfnamefont {M.}~\bibnamefont {David}}, \bibinfo
  {author} {\bibfnamefont {L.}~\bibnamefont {Martins}}, \bibinfo {author}
  {\bibfnamefont {C.}~\bibnamefont {Almeida}}, \bibinfo {author} {\bibfnamefont
  {E.}~\bibnamefont {Martins-Ferreira}}, \bibinfo {author} {\bibfnamefont
  {K.}~\bibnamefont {Takai}}, \bibinfo {author} {\bibfnamefont
  {T.}~\bibnamefont {Enoki}}, \bibinfo {author} {\bibfnamefont
  {R.}~\bibnamefont {Magalhães-Paniago}}, \bibinfo {author} {\bibfnamefont
  {A.}~\bibnamefont {Malachias}}, \bibinfo {author} {\bibfnamefont
  {A.}~\bibnamefont {Jorio}}, \bibinfo {author} {\bibfnamefont
  {B.}~\bibnamefont {Archanjo}}, \bibinfo {author} {\bibfnamefont
  {C.}~\bibnamefont {Achete}},\ and\ \bibinfo {author} {\bibfnamefont
  {L.}~\bibnamefont {Cançado}},\ }\bibfield  {title} {\enquote {\bibinfo
  {title} {Structural analysis of polycrystalline graphene systems by raman
  spectroscopy},}\ }\href
  {https://doi.org/https://doi.org/10.1016/j.carbon.2015.08.020} {\bibfield
  {journal} {\bibinfo  {journal} {Carbon}\ }\textbf {\bibinfo {volume} {95}},\
  \bibinfo {pages} {646--652} (\bibinfo {year} {2015})}\BibitemShut {NoStop}%
\bibitem [{\citenamefont {Inoue}\ \emph {et~al.}(1983)\citenamefont {Inoue},
  \citenamefont {Nakashima}, \citenamefont {Mitsuishi}, \citenamefont
  {Tabata},\ and\ \citenamefont {Tsuboi}}]{inoue}%
  \BibitemOpen
  \bibfield  {author} {\bibinfo {author} {\bibfnamefont {Y.}~\bibnamefont
  {Inoue}}, \bibinfo {author} {\bibfnamefont {S.}~\bibnamefont {Nakashima}},
  \bibinfo {author} {\bibfnamefont {A.}~\bibnamefont {Mitsuishi}}, \bibinfo
  {author} {\bibfnamefont {S.}~\bibnamefont {Tabata}},\ and\ \bibinfo {author}
  {\bibfnamefont {S.}~\bibnamefont {Tsuboi}},\ }\bibfield  {title} {\enquote
  {\bibinfo {title} {Raman spectra of amorphous sic},}\ }\href
  {https://doi.org/https://doi.org/10.1016/0038-1098(83)90834-7} {\bibfield
  {journal} {\bibinfo  {journal} {Solid State Communications}\ }\textbf
  {\bibinfo {volume} {48}},\ \bibinfo {pages} {1071--1075} (\bibinfo {year}
  {1983})}\BibitemShut {NoStop}%
\bibitem [{\citenamefont {Forbeaux}, \citenamefont {Themlin},\ and\
  \citenamefont {Debever}(1998)}]{Forbeaux}%
  \BibitemOpen
  \bibfield  {author} {\bibinfo {author} {\bibfnamefont {I.}~\bibnamefont
  {Forbeaux}}, \bibinfo {author} {\bibfnamefont {J.-M.}\ \bibnamefont
  {Themlin}},\ and\ \bibinfo {author} {\bibfnamefont {J.-M.}\ \bibnamefont
  {Debever}},\ }\bibfield  {title} {\enquote {\bibinfo {title} {Heteroepitaxial
  graphite on $6h\ensuremath{-}\mathrm{SiC}(0001):$ interface formation through
  conduction-band electronic structure},}\ }\href
  {https://doi.org/10.1103/PhysRevB.58.16396} {\bibfield  {journal} {\bibinfo
  {journal} {Phys. Rev. B}\ }\textbf {\bibinfo {volume} {58}},\ \bibinfo
  {pages} {16396--16406} (\bibinfo {year} {1998})}\BibitemShut {NoStop}%
\bibitem [{\citenamefont {Hass}\ \emph {et~al.}(2006)\citenamefont {Hass},
  \citenamefont {Feng}, \citenamefont {Li}, \citenamefont {Li}, \citenamefont
  {Zong}, \citenamefont {de~Heer}, \citenamefont {First}, \citenamefont
  {Conrad}, \citenamefont {Jeffrey},\ and\ \citenamefont {Berger}}]{Hass}%
  \BibitemOpen
  \bibfield  {author} {\bibinfo {author} {\bibfnamefont {J.}~\bibnamefont
  {Hass}}, \bibinfo {author} {\bibfnamefont {R.}~\bibnamefont {Feng}}, \bibinfo
  {author} {\bibfnamefont {T.}~\bibnamefont {Li}}, \bibinfo {author}
  {\bibfnamefont {X.}~\bibnamefont {Li}}, \bibinfo {author} {\bibfnamefont
  {Z.}~\bibnamefont {Zong}}, \bibinfo {author} {\bibfnamefont {W.~A.}\
  \bibnamefont {de~Heer}}, \bibinfo {author} {\bibfnamefont {P.~N.}\
  \bibnamefont {First}}, \bibinfo {author} {\bibfnamefont {E.~H.}\ \bibnamefont
  {Conrad}}, \bibinfo {author} {\bibfnamefont {C.~A.}\ \bibnamefont
  {Jeffrey}},\ and\ \bibinfo {author} {\bibfnamefont {C.}~\bibnamefont
  {Berger}},\ }\bibfield  {title} {\enquote {\bibinfo {title} {Highly ordered
  graphene for two dimensional electronics},}\ }\href
  {https://doi.org/10.1063/1.2358299} {\bibfield  {journal} {\bibinfo
  {journal} {Applied Physics Letters}\ }\textbf {\bibinfo {volume} {89}},\
  \bibinfo {pages} {143106} (\bibinfo {year} {2006})}\BibitemShut {NoStop}%
\bibitem [{\citenamefont {Berger}\ \emph {et~al.}(2006)\citenamefont {Berger},
  \citenamefont {Song}, \citenamefont {Li}, \citenamefont {Wu}, \citenamefont
  {Brown}, \citenamefont {Naud}, \citenamefont {Mayou}, \citenamefont {Li},
  \citenamefont {Hass}, \citenamefont {Marchenkov}, \citenamefont {Conrad},
  \citenamefont {First},\ and\ \citenamefont {de~Heer}}]{Berger}%
  \BibitemOpen
  \bibfield  {author} {\bibinfo {author} {\bibfnamefont {C.}~\bibnamefont
  {Berger}}, \bibinfo {author} {\bibfnamefont {Z.}~\bibnamefont {Song}},
  \bibinfo {author} {\bibfnamefont {X.}~\bibnamefont {Li}}, \bibinfo {author}
  {\bibfnamefont {X.}~\bibnamefont {Wu}}, \bibinfo {author} {\bibfnamefont
  {N.}~\bibnamefont {Brown}}, \bibinfo {author} {\bibfnamefont
  {C.}~\bibnamefont {Naud}}, \bibinfo {author} {\bibfnamefont {D.}~\bibnamefont
  {Mayou}}, \bibinfo {author} {\bibfnamefont {T.}~\bibnamefont {Li}}, \bibinfo
  {author} {\bibfnamefont {J.}~\bibnamefont {Hass}}, \bibinfo {author}
  {\bibfnamefont {A.~N.}\ \bibnamefont {Marchenkov}}, \bibinfo {author}
  {\bibfnamefont {E.~H.}\ \bibnamefont {Conrad}}, \bibinfo {author}
  {\bibfnamefont {P.~N.}\ \bibnamefont {First}},\ and\ \bibinfo {author}
  {\bibfnamefont {W.~A.}\ \bibnamefont {de~Heer}},\ }\bibfield  {title}
  {\enquote {\bibinfo {title} {Electronic confinement and coherence in
  patterned epitaxial graphene},}\ }\href
  {https://doi.org/10.1126/science.1125925} {\bibfield  {journal} {\bibinfo
  {journal} {Science}\ }\textbf {\bibinfo {volume} {312}},\ \bibinfo {pages}
  {1191--1196} (\bibinfo {year} {2006})}\BibitemShut {NoStop}%
\bibitem [{\citenamefont {Ouerghi}\ \emph {et~al.}(2010)\citenamefont
  {Ouerghi}, \citenamefont {Kahouli}, \citenamefont {Lucot}, \citenamefont
  {Portail}, \citenamefont {Travers}, \citenamefont {Gierak}, \citenamefont
  {Penuelas}, \citenamefont {Jegou}, \citenamefont {Shukla}, \citenamefont
  {Chassagne},\ and\ \citenamefont {Zielinski}}]{Ouerghi}%
  \BibitemOpen
  \bibfield  {author} {\bibinfo {author} {\bibfnamefont {A.}~\bibnamefont
  {Ouerghi}}, \bibinfo {author} {\bibfnamefont {A.}~\bibnamefont {Kahouli}},
  \bibinfo {author} {\bibfnamefont {D.}~\bibnamefont {Lucot}}, \bibinfo
  {author} {\bibfnamefont {M.}~\bibnamefont {Portail}}, \bibinfo {author}
  {\bibfnamefont {L.}~\bibnamefont {Travers}}, \bibinfo {author} {\bibfnamefont
  {J.}~\bibnamefont {Gierak}}, \bibinfo {author} {\bibfnamefont
  {J.}~\bibnamefont {Penuelas}}, \bibinfo {author} {\bibfnamefont
  {P.}~\bibnamefont {Jegou}}, \bibinfo {author} {\bibfnamefont
  {A.}~\bibnamefont {Shukla}}, \bibinfo {author} {\bibfnamefont
  {T.}~\bibnamefont {Chassagne}},\ and\ \bibinfo {author} {\bibfnamefont
  {M.}~\bibnamefont {Zielinski}},\ }\bibfield  {title} {\enquote {\bibinfo
  {title} {Epitaxial graphene on cubic sic(111)/si(111) substrate},}\ }\href
  {https://doi.org/10.1063/1.3427406} {\bibfield  {journal} {\bibinfo
  {journal} {Applied Physics Letters}\ }\textbf {\bibinfo {volume} {96}},\
  \bibinfo {pages} {191910} (\bibinfo {year} {2010})}\BibitemShut {NoStop}%
\bibitem [{\citenamefont {de~Heer}\ \emph {et~al.}(2011)\citenamefont
  {de~Heer}, \citenamefont {Berger}, \citenamefont {Ruan}, \citenamefont
  {Sprinkle}, \citenamefont {Li}, \citenamefont {Hu}, \citenamefont {Zhang},
  \citenamefont {Hankinson},\ and\ \citenamefont {Conrad}}]{deHeer}%
  \BibitemOpen
  \bibfield  {author} {\bibinfo {author} {\bibfnamefont {W.~A.}\ \bibnamefont
  {de~Heer}}, \bibinfo {author} {\bibfnamefont {C.}~\bibnamefont {Berger}},
  \bibinfo {author} {\bibfnamefont {M.}~\bibnamefont {Ruan}}, \bibinfo {author}
  {\bibfnamefont {M.}~\bibnamefont {Sprinkle}}, \bibinfo {author}
  {\bibfnamefont {X.}~\bibnamefont {Li}}, \bibinfo {author} {\bibfnamefont
  {Y.}~\bibnamefont {Hu}}, \bibinfo {author} {\bibfnamefont {B.}~\bibnamefont
  {Zhang}}, \bibinfo {author} {\bibfnamefont {J.}~\bibnamefont {Hankinson}},\
  and\ \bibinfo {author} {\bibfnamefont {E.}~\bibnamefont {Conrad}},\
  }\bibfield  {title} {\enquote {\bibinfo {title} {Large area and structured
  epitaxial graphene produced by confinement controlled sublimation of silicon
  carbide},}\ }\href {https://doi.org/10.1073/pnas.1105113108} {\bibfield
  {journal} {\bibinfo  {journal} {Proceedings of the National Academy of
  Sciences}\ }\textbf {\bibinfo {volume} {108}},\ \bibinfo {pages}
  {16900--16905} (\bibinfo {year} {2011})}\BibitemShut {NoStop}%
\bibitem [{\citenamefont {Giusca}\ \emph {et~al.}(2014)\citenamefont {Giusca},
  \citenamefont {Spencer}, \citenamefont {Shard}, \citenamefont {Yakimova},\
  and\ \citenamefont {Kazakova}}]{Giusca}%
  \BibitemOpen
  \bibfield  {author} {\bibinfo {author} {\bibfnamefont {C.~E.}\ \bibnamefont
  {Giusca}}, \bibinfo {author} {\bibfnamefont {S.~J.}\ \bibnamefont {Spencer}},
  \bibinfo {author} {\bibfnamefont {A.~G.}\ \bibnamefont {Shard}}, \bibinfo
  {author} {\bibfnamefont {R.}~\bibnamefont {Yakimova}},\ and\ \bibinfo
  {author} {\bibfnamefont {O.}~\bibnamefont {Kazakova}},\ }\bibfield  {title}
  {\enquote {\bibinfo {title} {Exploring graphene formation on the c-terminated
  face of sic by structural, chemical and electrical methods},}\ }\href
  {https://doi.org/https://doi.org/10.1016/j.carbon.2013.12.018} {\bibfield
  {journal} {\bibinfo  {journal} {Carbon}\ }\textbf {\bibinfo {volume} {69}},\
  \bibinfo {pages} {221--229} (\bibinfo {year} {2014})}\BibitemShut {NoStop}%
\bibitem [{\citenamefont {Luxmi}\ \emph {et~al.}(2010)\citenamefont {Luxmi},
  \citenamefont {Srivastava}, \citenamefont {He}, \citenamefont {Feenstra},\
  and\ \citenamefont {Fisher}}]{Luxmi}%
  \BibitemOpen
  \bibfield  {author} {\bibinfo {author} {\bibnamefont {Luxmi}}, \bibinfo
  {author} {\bibfnamefont {N.}~\bibnamefont {Srivastava}}, \bibinfo {author}
  {\bibfnamefont {G.}~\bibnamefont {He}}, \bibinfo {author} {\bibfnamefont
  {R.~M.}\ \bibnamefont {Feenstra}},\ and\ \bibinfo {author} {\bibfnamefont
  {P.~J.}\ \bibnamefont {Fisher}},\ }\bibfield  {title} {\enquote {\bibinfo
  {title} {Comparison of graphene formation on c-face and si-face sic {0001}
  surfaces},}\ }\href {https://doi.org/10.1103/PhysRevB.82.235406} {\bibfield
  {journal} {\bibinfo  {journal} {Phys. Rev. B}\ }\textbf {\bibinfo {volume}
  {82}},\ \bibinfo {pages} {235406} (\bibinfo {year} {2010})}\BibitemShut
  {NoStop}%
\bibitem [{\citenamefont {Leitgeb}\ \emph {et~al.}(2021)\citenamefont
  {Leitgeb}, \citenamefont {Pfusterschmied}, \citenamefont {Schwarz},
  \citenamefont {Depuydt}, \citenamefont {Cho},\ and\ \citenamefont
  {Schmid}}]{Leitgeb2021}%
  \BibitemOpen
  \bibfield  {author} {\bibinfo {author} {\bibfnamefont {M.}~\bibnamefont
  {Leitgeb}}, \bibinfo {author} {\bibfnamefont {G.}~\bibnamefont
  {Pfusterschmied}}, \bibinfo {author} {\bibfnamefont {S.}~\bibnamefont
  {Schwarz}}, \bibinfo {author} {\bibfnamefont {B.}~\bibnamefont {Depuydt}},
  \bibinfo {author} {\bibfnamefont {J.}~\bibnamefont {Cho}},\ and\ \bibinfo
  {author} {\bibfnamefont {U.}~\bibnamefont {Schmid}},\ }\bibfield  {title}
  {\enquote {\bibinfo {title} {Communication{\textemdash}current oscillations
  in photoelectrochemical etching of monocrystalline 4h silicon carbide},}\
  }\href {https://doi.org/10.1149/2162-8777/ac10b3} {\bibfield  {journal}
  {\bibinfo  {journal} {{ECS} Journal of Solid State Science and Technology}\
  }\textbf {\bibinfo {volume} {10}},\ \bibinfo {pages} {073003} (\bibinfo
  {year} {2021})}\BibitemShut {NoStop}%
\end{thebibliography}%

\end{document}